\documentclass[11pt,a4paper]{article}
\pdfoutput=1
\usepackage{amsmath}
\usepackage{amssymb}
\usepackage{theorem}
\usepackage{array}
\usepackage{graphicx}
\usepackage{subfig}
\usepackage[numbers, sort]{natbib}
\usepackage{hyperref}

\usepackage[toc, page]{appendix}

\numberwithin{equation}{section}

\begin{document}

\title{\huge \bf{Deviations from Tri-bimaximal Mixing: Charged Lepton Corrections and
Renormalization Group Running}}
\author{
    S.Boudjemaa\footnote{E-mail:sally83@phys.soton.ac.uk},
    S.F.King\footnote{E-mail: sfk@hep.phys.soton.ac.uk}\\
    School of Physics and Astronomy, University of Southampton,\\
    Southampton, SO17 1BJ, U.K}
\maketitle

\begin{abstract}
\noindent
We analyze the effects of charged lepton corrections and
renormalization group (RG) running on the low energy predictions of
theories which accurately predict
tri-bimaximal neutrino mixing at the high energy scale.
In particular we focus on GUT inspired see-saw models
with accurate tri-bimaximal neutrino mixing at the GUT scale,
in which the charged lepton corrections
are Cabibbo-like and give rise to sum rules valid at the GUT scale.
We study numerically the RG corrections to a variety of such neutrino mixing sum rules
in order to assess their accuracy and reliability when comparing them to
future low energy neutrino oscillation experiments.
Our results indicate that the RG corrections to neutrino mixing sum rules are
typically small (less than one degree), at least in the examples studied with hierarchical neutrinos.
\end{abstract}

\section{Introduction}

Perhaps the greatest advance in particle physics
over the past decade has been the discovery of
neutrino mass and mixing involving two large mixing angles commonly known as the
atmospheric angle $\theta_{23}$ and the solar angle $\theta_{12}$. The latest
data from neutrino oscillation experiments is consistent with the so
called tri-bimaximal (TB) mixing pattern \cite{tribi},
\begin{equation}
\label{TBM}
U_{TB}= \left(\begin{array}{ccc} \sqrt{\frac{2}{3}}& \frac{1}{\sqrt{3}}&0\\
-\frac{1}{\sqrt{6}}&\frac{1}{\sqrt{3}}&\frac{1}{\sqrt{2}}\\
\frac{1}{\sqrt{6}}&-\frac{1}{\sqrt{3}}&\frac{1}{\sqrt{2}} \end{array} \right)
P_{Maj},
\end{equation}
where $P_{Maj}$ is the diagonal phase matrix involving the two observable
Majorana phases, and there were many attempts to reproduce this as a theoretical
prediction \cite{King:2005bj,Frampton:2004ud,Altarelli:2006kg,Ma:2007wu,
deMedeirosVarzielas:2005ax,King:2006np,
Harrison:2003aw,Chan:2007ng,Antusch:2007jd}. Since the forthcoming
neutrino experiments will be
sensitive to small deviations from TB mixing, it is important to study the
theoretical uncertainty in such TB mixing predictions.

The question of how to achieve TM mixing has been the subject of intense
theoretical speculation. In theoretical models one attempts to
construct the neutrino and charged lepton mass matrices in some particular basis.
There are two particular bases that have been used widely in the literature for this
purpose, as follows. The first basis is
the flavour basis in which the charged lepton mass matrix is diagonal,
while the neutrino mass matrix takes a particular form such that is results
in TB mixing. The second basis is a particular basis first introduced by Cabibbo
and Wolfenstein in which both the neutrino and charged lepton mass matrices
are non-diagonal, but in which the
charged lepton mass matrix is diagonalised by a ``democratic unitary matrix''
involving elements of equal magnitude but differing by a phase $\omega=2\pi/3$.
Such a Cabibbo-Wolfenstein basis is particularly well suited to models of TB mixing
based on the discrete group $A_4$ \cite{Ma}. However in other classes of models,
one attempts to work in the flavour basis and to derive TB mixing purely from the
neutrino sector with the charged lepton matrix being diagonal, for example
using constrained sequential dominance (CSD) \cite{King:2005bj}.

However, when attempting to derive TB mixing in the flavour basis
in realistic models arising from Grand Unified Theories (GUTs), it
is observed that, although TB mixing may be accurately achieved
from neutrino mixing, in practice the flavour basis is never
accurately achieved, i.e. the charged lepton mass matrix is never
accurately diagonal. Instead, in such GUT models, the charged
lepton mass matrix often resembles the down quark mass matrix, and
involves an additional Cabibbo-like rotation in order to
diagonalize it. In such models, then, TB mixing arises in the
neutrino sector, but with charged lepton correction giving
deviations \cite{Xing:2002sw}. It turns out that such Cabibbo-like
charged lepton corrections lead to well defined corrections to TB
mixing which can be cast in the form of sum rules expressed in
terms of the measurable PMNS parameters. One example is the
neutrino mixing sum rule $\theta_{12} -\theta_{13} \cos (\delta)
\approx
\theta_{12}^\nu$\cite{King:2005bj,Masina:2005hf,Antusch:2005kw},
where $\delta$ is the observable Dirac CP phase in the standard
parameterisation. Since such sum rules may be tested in future
high precision long baseline neutrino experiments, it is of
interest to know with what precision they are expected to hold
theoretically.

Such sum rules as discussed above are strictly only expected to apply
at some high energy scale, whereas the neutrino experiments are
performed at low energy scales.  In order to compare the predictions
of such sum rules to experiment one must therefore perform a
renormalisation group (RG) running from the high energy (e.g. the
Grand Unified Theory (GUT)) scale where the theory is defined to the
electroweak scale $M_{Z}$. RG corrections arise mainly from the large
tau lepton and third family neutrino Yukawa couplings, and this leads
to large wavefunction corrections in the framework of supersymmetric
models. The running of neutrino masses and lepton mixing angles is
very important and has been studied extensively in
the literature \cite{Chankowski:1993tx}. In
\cite{Antusch:2002hy,Antusch:2005gp} a mathematica package
REAP (http://www.ph.tum.de/~rge/) which solves RGEs and provides
numerical values for the neutrino mass matrix and mixing angles was
developed.

In this paper we provide a first numerical study of the deviations from TB mixing
due to the effects of both charged lepton
corrections and RG running. We focus on GUT inspired models in which the charged lepton
corrections
are Cabibbo-like and in this case they may be cast in terms of sum rules valid at the GUT scale.
In practice, then, we are interested in the RG corrections to these sum rules which may
subsequently
be reliably compared to experiment at low energies. We shall study
a variety of neutrino mixing sum rules (arising from the deviations from exact
tri-bimaximal neutrino mixing due to Cabibbo-like charged lepton corrections)
and comment on their accuracy and reliability when comparing them to
future low energy neutrino oscillation experiments.
Most of the specific numerical results are inspired by a particular class of
GUT-flavour models, namely the models in
\cite{King:2005bj,deMedeirosVarzielas:2005ax}, which are precisely the type
of models in which the sum rules emerge in the first place.
In the cases studied we find rather small corrections.
For example the sum rule
$\theta_{12} -\theta_{13} \cos (\delta) \approx 35.3^o$
becomes renormalized by about $0.4^o$
for large $\tan \beta=50$.
Although most of the numerical results are based on a particular GUT motivated model,
we also analyze a completely different type of model
and find qualitatively
similar results. This suggests that our results will apply to more general models
based on the Minimal Supersymmetric
Standard Model, extended to include the see-saw mechanism,
with hierarchical neutrino masses.

We emphasize that in this paper
we only consider deviations from TB lepton mixing due to the
combination of charged lepton corrections
and RG running, and that in general there will be other sources of deviations
that we do not consider. For example, exact CSD \cite{King:2005bj} itself will lead to some
deviations since it does not predict precisely TB neutrino mixing
due to corrections of order $m_2/m_3$ \cite{King:1998jw}.
Another example of deviations to TB mixing are the
canonical normalization effects discussed in \cite{Antusch:2007ib}.
Clearly such additional corrections are quite model dependent and
a phenomenological study of corrections to the TB mixing in the
neutrino sector, in the flavour basis, has been made
in \cite{Albright}. In this paper we shall simply assume precise TB neutrino mixing at the
GUT scale, and investigate the deviations due to charged lepton corrections
and RG running only, ignoring other possible model dependent corrections.
The RG corrections to TB mixing (but not charged lepton corrections
or the resulting sum rules) were also considered in \cite{Dighe:2006sr}.

The layout of the remainder of the paper is as follows.
In section 2 we give our conventions, including the TB deviation
parameters that we will use. In section 3 we discuss the see-saw
mechanism and show how TB neutrino mixing can be achieved.
In section 4 we show how Cabibbo-like charged lepton
corrections leads to neutrino mixing sum rules.
We also present a numerical model of TB neutrino mixing
adapted from a well motivated example of light sequential
dominance used in the GUT flavour models of \cite{King:2005bj,deMedeirosVarzielas:2005ax}
that we shall use in most of the remainder of the paper. We also
show how sensitive these results are to non-Cabibbo like
charged lepton corrections. In section 5 we numerically study the
RG corrections to the various neutrino mixing sum rules
which embody the charged lepton corrections to TB neutrino mixing.
In section 6 we explore a second type of numerical model
adapted from heavy sequential dominance \cite{King:1998jw} and show that
the results are qualitatively similar to the case of the first numerical model.
Section 7 concludes the paper.

\section{Conventions}\label{conv}

\subsection{The PMNS matrix in the standard parametrization}

The mixing matrix in the lepton sector, the PMNS matrix $U_{PMNS}$, is defined
as the matrix appearing in the electroweak coupling to the $W$ bosons
expressed
in terms of lepton mass eigenstates. The Lagrangian is given in terms of mass
matrices of charged leptons $M_e$ and neutrinos $m_\nu$ as,
\begin{equation}
\mathcal{L}=-\bar{e}_L M_e e_R -\frac{1}{2} \bar{\nu}_L m_{LL}\nu_L^c + H.c,
\end{equation}

The change in basis from flavour to eigenbasis is performed by,
\begin{equation}
V_{e_L} M_e V_{e_R}^\dagger = diag(m_e, m_\mu, m_\tau), \hspace{3mm}
V_{\nu_L}
m_{LL} V_{\nu_L}^T =diag(m_1, m_2, m_3),
\end{equation}
The PMNS matrix is then given by,
\begin{equation}
U_{PMNS} =V_{e_L} V_{\nu_L}^\dagger.
\label{PMNS}
\end{equation}

In the standard PDG parametrization, the PMNS matrix can be written as,

\begin{equation}
\label{MNS}
U_{PMNS}=\left( \begin{array}{ccc} c_{12} c_{13} & s_{12} c_{13} & s_{13} e^{- i
\delta}\\ -c_{23} s_{12} -s_{13} s_{23} c_{12} e^{i \delta} & c_{23} c_{12} -
s_{13} s_{23} s_{12} e^{i \delta} & s_{23} c_{13}\\ s_{23} s_{12} - s_{13}
c_{23} c_{12} e^{i \delta} & -s_{23} c_{12} -s_{13} c_{23} s_{12} e^{i \delta} &
c_{23} c_{13} \end{array}\right) P_{Maj},
\end{equation}
where $\delta$ is the Dirac CP violating phase, and $P_{Maj}=diag(e^{i
\frac{\alpha_1}{2}}, e^{i \frac{\alpha_2}{2}}, 0)$ contains the Majorana phases
$\alpha_1$, $\alpha_2$.  The latest experimental values and errors for the three
neutrino oscillation parameters are summarised inTable.\ref{tab12}
\cite{Schwetz:2007my, Maltoni:2003da, Maltoni:2004ei}.

\begin{table}[hbtp]
    \centering
    \begin{tabular}{ c  c  c  c }
    \hline
    \\[-10pt]
    Parameter & Best fit ($\,^{\circ}$)& 2 $\sigma$ ($\,^{\circ}$)& 3 $\sigma$ ($
\,^{\circ}$) \\
    \\[-10pt]
     \hline
    \\[-10pt]

    $\theta_{12}$ & 34.44   & 31.94- 37.46 & 30.65- 39.23   \\

    $\theta_{23}$ & 45  & 38.05 - 52.53 & 35.66 - 54.93 \\

     $\theta_{13}$ & 4.79 & $\leq$  10.46 & $\leq$ 12.92 \\
     \\[-10pt]
     \hline
  \end{tabular}

    \caption{ Best fit values, 2 $\sigma$ and 3 $\sigma$ intervals for the three-
flavour neutrino oscillation parameters from global data including accelerator
(K2K and MINOS) and solar, atmospheric, reactor (Kam LAND and CHOOZ)
experiments \cite{Schwetz:2007my}.}

    \label{tab12}

\end{table}

\subsection{A parametrization in terms of deviations}

Another parametrisation of the lepton mixing matrix can be achieved by taking an
expansion about the tri-bimaximal matrix. Three small parameters $r$, $s$ and
$a
$ are introduced to describe the deviations of the reactor, solar and atmospheric
angles from their tri-bimaximal values \cite{King:2007pr},

\begin{equation}\label{de}
s_{13}= \frac{r}{\sqrt{2}}, \hspace{4mm}  s_{12} = \frac{1}{\sqrt{3}} (1+ s),
\hspace{4mm} s_{23} = \frac{1}{\sqrt{2}} (1+ a).
\end{equation}

Global fits of the conventional mixing angles \cite{Maltoni:2004ei} can be
translated into the ranges,
\begin{equation}\label{bf}
0 < r  <  0.22, \hspace{4mm} -0.11 < s < 0.04, \hspace{4mm} -0.12 < a < 0.13.
\end{equation}

Considering an expansion of the lepton mixing matrix in powers of $r$, $s$, $a$
about the tri-bimaximal form. One gets the following form for the mixing matrix to
first order in  $r$, $s$, $a$ \cite{King:2007pr},

\begin{equation}\label{Dev}
U_{MNS} \approx \left( \begin{array}{ccc} \sqrt{\frac{2}{3}} (1-\frac{1}{2} s) &
\frac{1}
{\sqrt{3}} (1+s) & \frac{1}{\sqrt{2}} r e^{- i \delta}\\
-\frac{1}{\sqrt{6}} (1+ s - a + r e^{i \delta}) & \frac{1}{\sqrt{3}} (1 - \frac{1}{2} s - a -
\frac{1}{2} r e^{i \delta}) & \frac{1}{\sqrt{2}} (1 + a)\\
\frac{1}{\sqrt{6}} (1 + s + a - r e^{i \delta}) & -\frac{1}{\sqrt{3}} (1 - \frac{1}{2} s + a +
\frac{1}{2} r e^{i \delta}) & \frac{1}{\sqrt{2}} (1 - a)
 \end{array} \right) P_{Maj}.
\end{equation}

\subsection{Third row deviation parameters}
For later convenience, we also define the following parameters which
express the deviation of the magnitude of the third row mixing matrix
elements from
their tri-bimaximal values:
\begin{eqnarray}\label{xi1}
|(U_{PMNS})_{31}| &\equiv  & \frac{1}{\sqrt{6}}(1+\xi_1 )\nonumber \\
|(U_{PMNS})_{32}| &\equiv  & \frac{1}{\sqrt{3}}(1+\xi_2 )\nonumber \\
|(U_{PMNS})_{33}| &\equiv  & \frac{1}{\sqrt{2}}(1+\xi_3 )
\end{eqnarray}

Hence from Eq.\ref{MNS},
\begin{equation}\label{xi2}
\begin{array}{ccc}
\xi_1 & = & \sqrt{6}~ |s_{23} s_{12} -
s_{13} c_{23} c_{12} e^{i \delta}| - 1 , \\
\xi_2 & = & \sqrt{3}~ |-s_{23} c_{12} -
s_{13} c_{23} s_{12} e^{i \delta} | - 1 ,\\
\xi_3 & = & \sqrt{2}~ |c_{23}
c_{13}|-1.
 \end {array}
 \end{equation}

We can also express the $\xi_i$ parameters in terms of the
deviation parameters $r$, $s$, $a$ from Eq.\ref{Dev}as follows,
\begin{equation}\label{xi3}
\begin{array}{ccc}
\xi_1 & \approx &| 1+ s + a - r e^{i \delta}| -1,\\
\xi_2 & \approx & |1 - \frac{1}{2} s + a + \frac{1}{2} r e^{i \delta}| -1,\\
\xi_3 & \approx &|1- a| - 1.
\end{array}
\end{equation}

\section{The See-Saw Mechanism and TB Neutrino Mixing}\label{one}

\subsection{The see-saw mechanism}
The see-saw mechanism provides an excellent explanation for the smallness of
neutrino masses.
Before discussing its simplest form, we first start by summarising the possible
types of neutrino mass.

One type is Majorana masses of the form $m_{LL}\bar{\nu_{L}} \nu_L^c$ where $
\nu_L$ is a left-handed neutrino field and $\nu_L^c$ is the CP conjugate of a left-
handed neutrino field, in other words a right handed antineutrino field \cite{King:2007nw}.
Introducing right-handed neutrino fields, other neutrino mass terms are possible.
There are additional Majorana masses of the form $M_{RR}\bar{\nu_R} \nu_R^c
$
and Dirac masses of the form $m_{LR}\bar{\nu_L} \nu_R$ where $\nu_R$ is a
right-handed neutrino field and $\nu_R^c$ is its CP conjugate.

The Majorana masses of the form $ m_{LL}^{\nu} $are strictly forbidden in the
standard model, assuming only the higgs doublets are present. The reason for
this is that heavy left-handed neutrinos would disturb the theory of weak
interactions with W, Z bosons. For the simplest version of the see-saw
mechanism,
one can assume that the left-handed Majorana masses are zero at first, but are
effectively generated after introducing the right handed neutrinos
 \cite {seesaw}.

The right-handed neutrino does not take part in weak interactions with the W, Z
bosons, and so its mass $M_{RR}$ can be arbitrarily large. With these types of
neutrino mass, the see-saw mass matrix is given as,
\begin{equation}\label{m}
\left( \bar{\nu_L } \bar{\nu_R^c} \right) \left( \begin{array}{cc} 0& m_{LR}\\
m_{LR}^T&M_{RR}  \end{array} \right)\left( \begin{array}{cc} \nu_L^c\\ \nu_R
\end{array}\right)
\end{equation}

In the approximation that $M_{RR} \gg m_{LR}$ ( $M_{RR}$ may be orders of
magnitude larger than the electroweak scale), the matrix in Eq.\ref{m} can be
diagonalised to give the effective left-handed Majorana masses,
\begin{equation}\label{see}
m_{LL}=m_{LR} M_{RR}^{-1} m_{LR}^T
\end{equation}

These masses are naturally suppressed by the heavy scale $M_{RR}$. Taking
$m_{LR}=M_W=80 GeV$ and $M_{RR}=M_{GUT}=10^{16} GeV$, we find
$m_{LL} \sim 10^{-3} eV$ which is good for solar neutrinos. A right-handed
neutrino with a mass below the GUT scale would be required for atmospheric
neutrinos.

The fundamental parameters which must be inputted into the see-saw
mechanism
are the Dirac mass matrix $m_{LR}$ and the heavy right-handed neutrino
Majorana mass matrix $M_{RR}$. The output is the effective left-handed
Majorana
mass matrix $m_{LL}$ as given by the see-saw formula in Eq.\ref{see}
\cite{Antusch:2004gf}.

The see-saw mechanism discussed so far is the simplest version and it is
sometimes called type I see-saw mechanism. In Pati-Salam models or grand
unified theories based on $SO(10)$, type I is generalised to type II see-saw,
where an additional term for the light neutrinos is present \cite{lazarides:1981}.

\subsection{Approximate TB neutrino mixing from CSD}

Sequential dominance (SD) is an elegant way of accounting for a neutrino mass
hierarchy with large atmospheric and solar mixing angles
\cite{King:2002nf, King:2002qh}. Here we review how
tri-bimaximal neutrino mixing can result from
constrained sequential dominance (CSD) \cite{King:2005bj}.
In SD, the atmospheric and solar neutrino mixing angles
are obtained in terms of ratios of Yukawa couplings involving the dominant and
subdominant right-handed neutrinos, respectively. To understand how tri-
bimaximal neutrino mixing could emerge from SD, we begin by writing the right-
handed neutrino Majorana mass matrix $M_{RR}$ in a diagonal basis as,
\begin{equation}
M_{RR} \approx \left(\begin{array}{ccc} Y&0&0\\ 0&X&0\\ 0&0&X' \end{array}
\right).
\end{equation}
Without loss of generality write the neutrino (Dirac) Yukawa matrix $Y_{LR}^\nu$ in terms
of the complex Yukawa couplings a,b,c,d,e,f,a',b',c' as
\begin{equation}
\label{abc}
Y_{LR}^\nu =\left(\begin{array}{ccc} d&a&a' \\ e&b&b' \\ f&c&c' \end{array} \right).
\end{equation}

For simplicity we assume that $d=0$. SD then corresponds to the
right-handed neutrino of mass $Y$ being the dominant term while the right-
handed neutrino of mass $X$ giving the leading sub-dominant contribution to the
see-saw mechanism.
\begin{equation}\label{y}
\frac{|e^2|, |f^2|, |ef|}{Y} \gg \frac{|xy|}{X} \gg \frac{x' y'}{X'}
\end{equation}
where $x, y \in a, b, c $ and $ x', y' \in a', b' c' $, and all Yukawa couplings are
assumed to be complex. Light sequential dominance corresponds to
\begin{equation}\label{Y}
Y < X < X' .
\end{equation}

Tri-bimaximal neutrino mixing, in which $\tan \theta_{23}^\nu =1$, $\tan
\theta_{12}^\nu =1/ \sqrt{2}$ and $\theta_{13}^\nu= 0 $ corresponds to the
choice,
\begin{eqnarray}
\label{CSD}
|a|&=&|b|=|c|,\\
|d|&=&0,\\
|e|&=&|f|\\
e^*b + f^* c&=&0.
\end{eqnarray}
This corresponds to
constrained sequential dominance (CSD)\cite{King:2005bj}.
Note that the analytic results for SD and CSD are accurate to
leading order in $m_2/m_3$ \cite{King:1998jw}, so these conditions will not give rise
precisely to TB neutrino mixing, and so in the numerical studies
we shall need to perturb the CSD relations in order to achieve
accurate TB neutrino mixing at the GUT scale, as discussed later.

\section{Charged lepton corrections}\label{ds}

\subsection{Cabibbo-like corrections}

\subsubsection{Sum rules}
In this paper we shall consider the case that TB mixing (Eq.\ref{TBM}) applies
quite accurately only to the {\em neutrino mixing} in some basis where the
charged lepton mass matrix is not exactly diagonal \cite{ChargedLeptonCorrections}.
This is a situation often
encountered in realistic models \cite{King:2005bj}.
Furthermore in GUT models it is often the case that, in the basis where the neutrino
mixing is of the TB form,
the charged lepton mixing matrix has a Cabibbo-like structure
rather similar to the quark mixing and
is dominated by a 1-2 mixing $\theta_{12}^E$ \cite{Antusch:2007dj},
\begin{equation}\label{cl}
V_{e_L} =  \left( \begin{array}{ccc} c_{\theta_{12}^E} & -s_{\theta_{12}^E}
e^{- i  \lambda_{12}^E} &0\\
s_{\theta_{12}^E}  e^{i \lambda_{12}^E} &c_{\theta_{12}^E} &0\\
0 & 0& 1
 \end{array} \right),
\end{equation}
where $c_{\theta_{12}^E} \equiv \cos \theta_{12}^E$, $s_{\theta_{12}^E} \equiv
\sin \theta_{12}^E$, and $\lambda_{12}^E$ is a phase required for the
diagonalisation of the charged lepton mass matrix \cite{King:2005bj}.
The physical PMNS oscillation
phase $\delta$ turns out to be
related to $ \lambda_{12}^E$ by \cite{Antusch:2007dj},
\begin{equation}
\delta=\lambda_{12}^E +\pi.
\end{equation}
In this paper we assume that
the neutrino mixing is accurately of the TB form,
\begin{equation}
\label{TBMnu}
V_{\nu_L}^{\dagger}= \left(\begin{array}{ccc} \sqrt{\frac{2}{3}}& \frac{1}{\sqrt{3}}&0\\
-\frac{1}{\sqrt{6}}&\frac{1}{\sqrt{3}}&\frac{1}{\sqrt{2}}\\
\frac{1}{\sqrt{6}}&-\frac{1}{\sqrt{3}}&\frac{1}{\sqrt{2}} \end{array} \right)
P_{Maj}.
\end{equation}
The physical mixing matrix is given by
Eq.\ref{PMNS} using Eq.\ref{TBMnu} and Eq.\ref{cl}.
The standard PDG form of the PMNS mixing matrix in Eq.\ref{MNS}
requires real elements $(U_{PMNS})_{11}$ and $(U_{PMNS})_{12}$ and this may be
achieved by use of the phases in
$P_{Maj}=diag(e^{i
\frac{\alpha_1}{2}}, e^{i \frac{\alpha_2}{2}}, 0)$.

It follows that $(U_{PMNS})_{31}$, $
(U_{PMNS})_{32}$ and $(U_{PMNS})_{33}$ are unaffected by
the Cabibbo-like charged lepton corrections and are hence
given by:
\begin{equation}\label{sum}
|(U_{PMNS})_{31}| = |(V_{\nu_L}^\dagger)_{31}| = \frac{1}{\sqrt{6}},
\end{equation}
\begin{equation}\label{sumtt}
|(U_{PMNS})_{32}|= |(V_{\nu_L}^\dagger)_{32}| = \frac{1}{\sqrt{3}},
\end{equation}
\begin{equation}\label{sumh}
|(U_{PMNS})_{33}|=|(V_{\nu_L}^\dagger)_{33}|=\frac{1}{\sqrt{2}}.
\end{equation}
The relations in Eqs.\ref{sum}, \ref{sumtt},
\ref{sumh} may be expressed
in terms of the third family matrix element deviation parameters
defined in Eq.\ref{xi1} as simply:
\begin{equation}
\label{xi0}
\xi_i =0.
 \end{equation}

Since these relations are all on the same footing, it is sufficient to
discuss one of them only and in the following we choose to focus on
Eq.~\ref{sum}. From Eq.\ref{MNS}, Eq.~\ref{sum} may be expanded
in terms of the standard mixing angles and phase leading to the
so called mixing sum rules as follows:
\begin{equation}\label{sm}
\Gamma_1 \equiv \arcsin \left(\sqrt{2}~|s_{23} s_{12}- s_{13} c_{23} c_{12} e^{i
\delta}|\right) = 35.26^o,
\end{equation}
where we have assumed $s_{23}^\nu \equiv \sin \theta_{23}^\nu = \frac{1}{\sqrt{2}}$.
This sum rule can be simplified further to leading order in $s_{13}$,
\begin{equation}\label{su}
\Gamma_2 \equiv \arcsin \left(\sqrt{2}~(s_{23} s_{12}- s_{13} c_{23} c_{12} \cos
{ \delta})\right) \approx 35.26^o.
\end{equation}
From Eq.\ref{sumh} $s_{23}=c_{23}=1/\sqrt{2}$, hence to leading order,
\begin{equation}
\label{gamma3}
\Gamma_3 \equiv  \theta_{12} - \theta_{13} \cos (\delta) \approx 35.26^o.
\end{equation}
The last form of the sum rule was first presented in
\cite{King:2005bj},
while all the forms can be found in \cite{Antusch:2005kw}.
We shall later study all three forms of the sum rules $\Gamma_i$,
together with some related sum rules which we now discuss.

Using the parametrization in Eq.\ref{de},
the sum rule in Eq.\ref{gamma3}
can be expressed in terms of the deviation parameters $s$, $r$ and the
Dirac CP phase ($\delta$)\cite{King:2007pr},
\begin{equation}\label{sd1}
\sigma_1 = r \cos \delta - s = 0.
\end{equation}
To deal with issues of canonical normalisation
corrections, the following sum rule has been proposed
\cite{Antusch:2007ib},
\begin{equation}\label{sd2}
\sigma_2 =  r \cos \delta + \frac{2}{3} a - s = 0.
\end{equation}
This sum rule was claimed to be stable under
leading logarithmic third family RG corrections, although,
as emphasized in \cite{Antusch:2007ib}, it does not include the
effect of the running of $\theta_{13}$ or $r$, whose inclusion introduces
a Majorana phase dependence.
\footnote{This sum rule was derived from an expansion in $m_2/m_3$,
and the running of $r$ was neglected because it is suppressed by an extra factor of $m_2/m_3$
compared to the running of $s$ and $a$.} Such effects will be studied numerically later.

\subsubsection{A GUT-Flavour Inspired Numerical Example}\label{lsd}
In order to study the RG corrections and reliability of the
various sum rules numerically it is necessary to define the
GUT scale matrices rather specifically.
In most of this paper we shall consider a numerical example inspired
by the GUT-flavour models of \cite{King:2005bj,deMedeirosVarzielas:2005ax},
although in Section 6 another numerical model will be considered
leading to qualitatively similar results.
Therefore in most of the remainder of this paper we shall
take the right-handed neutrino Majorana mass matrix $M_{RR}$
to be the diagonal matrix:
\[
M_{RR}=\left( \begin{array}{ccc} 5.1 \times 10^{-9} & 0 & 0 \\
0 & 7.05 \times 10^{-9} & 0 \\
0 & 0 & 1 \end{array} \right) M_3,
\]
where $M_3 = 10^{16} GeV$.
This is an example with light sequential dominance where the lightest right handed neutrino
is dominant \cite{King:1998jw}. Ignoring RGE corrections to begin with,
we find that precise tri-bimaximal neutrino mixing ($\theta^{\nu}_{12}= 35.26  \,^{\circ}$, $
\theta^{\nu}_{23} =45.00  \,^{\circ}$, $\theta^{\nu}_{13} =0.00  \,^{\circ}$)
can be achieved with the Yukawa matrix:
\begin{equation}\label{cor}
Y_{LR}^\nu =\left(\begin{array}{ccc} 0& 1.061667 b & 0.001\\
e  & b  & 0\\
- 0.9799 e  & b  & c_3\end{array} \right)
\end{equation}
where $b= 8.125 \times 10^{-5}$, $e= 2.125 \times 10^{-4}$
and $c_3= 0.5809$.
These parameters also lead to
the following values for the neutrino
masses: $m_1 = 1.75 \times 10^{-4} eV$, $m_2 = 8.67 \times 10^{-3} eV$,
$m_3 = 4.95 \times 10^{-2} eV$, $\Delta m^2_{atm} = 2.37 \times 10^{-3} eV^2$ and
 $\Delta m^2_{sol} = 7.52 \times 10^{-5} eV^2$.

The low energy pole masses of the quarks are given as follows: $m_u = 1.22~ MeV$,
$m_d = 2.77~MeV$, $m_s = 53 ~MeV$, $m_c = 0.595~GeV $, $m_b = 2.75~ GeV$
and $m_t = 163.6 ~GeV$. In order to satisfy these values at low energy scale, REAP was
used to perform the running of these masses from the $M_Z$ scale to the GUT scale and
the resulting quark Yukawa matrices $Y_u$ and $Y_d$ at the GUT scale were taken as
initial conditions for the running of the neutrino mixing parameters and sum rules from
 the GUT scale to $M_Z$ scale.

The above parameter choice approximately satisfies the CSD conditions in
Eq.\ref{CSD}. However small corrections are used in order to achieve
TBM neutrino mixing angles to 2 decimal places. If the CSD conditions
were imposed exactly we would find instead
$\theta_{12}= 33.97  \,^{\circ}$, $\theta_{23} =44.38  \,^{\circ}$, $\theta_{13} =0.059
\,^{\circ}$ and $\delta = 0  \,^{\circ}$ which are close to, but not accurately equal
to, the TBM values. This is to be expected since the SD relations are only
accurate to leading order in $m_2/m_3$ \cite{King:1998jw}. Since in this paper
we are interested in studying the deviations from exact TB neutrino mixing due to
charged lepton corrections and RG running we shall assume the matrices in
Eq.\ref{cor} rather than the CSD conditions as the starting point for our analysis.

In order to study the effect of Cabibbo-like charged lepton corrections
on the physical mixing angles where the neutrino
mixing is precisely tri-bimaximal, we shall use the REAP package previously
discussed. In order to use the REAP package it is convenient to work in the basis where
the charged lepton Yukawa matrix is diagonal. Thus,
assuming charged lepton corrections of the form of Eq.\ref{cl}, the neutrino
Yukawa matrix in the non-diagonal charged lepton basis must be transformed
to the diagonal charged lepton basis according to:
\begin{equation}
\label{rot}
Y_\nu \rightarrow Y'_\nu = V_{e_L} Y_\nu.
\end{equation}
Hence the original neutrino Yukawa matrix in Eq.\ref{cor}
must be rotated to the diagonal charged lepton basis according to
Eq.\ref{rot}.

Including the Cabibbo-like charged lepton corrections, physical tri-bimaximal
mixing only holds
when $ \theta_{12}^E = 0$.
However according to the sum rules for $\Gamma_i$,
certain combinatioms of mixing parameters sum to $35.262 \,^{\circ} $
for all values of the Cabibbo-like charged lepton corrections.
This is illustrated in Tables.\ref{tab1} ,\ref{tab2} where
the values of the mixing angles together with the
Dirac phase and the sum rules $\Gamma_1$, $\Gamma_2$, $\Gamma_3$ at
the GUT scale are presented for different values of
$\theta_{12}^E$ and $\lambda_{12}^E$. $\Gamma_1$ was found to be the most
accurate sum rule at the GUT scale with a value of $35.262 \,^{\circ} $ exactly at
all values of $ \theta_{12}^E$ and $ \lambda_{12}^E$. However the error in all the
sum rules is less than about $0.1^{\circ} $ in all the examples considered.

\begin{table}[hbtp]
    \centering

    \begin{tabular}{|l|c|c|c|c|c|}
    \hline

    $\theta_{12}^E$ & 0 & 1& 3 & 5 & 8  \\
    \hline

    $\theta_{12} $ & 35.26   & 34.648  & 33.429 & 32.216 & 30.407  \\
    \hline

   $\theta_{13}$ & 0.001 & 0.708 & 2.122 & 3.534 & 5.648\\
   \hline

   $\theta_{23}$ & 45.001& 44.997 & 44.962 & 44.892& 44.721\\
     \hline

    $\delta $& 0 & 210.204& 210.82 & 211.492& 212.672\\
    \hline

    $\Gamma_1$&35.262 & 35.262& 35.262& 35.262&35.262 \\
     \hline

   $\Gamma_2$& 35.262&35.26 &35.247 &35.217 &35.133 \\
    \hline

   $\Gamma_3$&35.261 &35.26 &35.252 &35.23 &35.162 \\
    \hline

\end{tabular}

\caption{Values of the neutrino mixing angles $\theta_{12}$, $\theta_{13}$ and $
\theta_{23}$ together with $\delta$ and the sum rules $\Gamma_1$, $
\Gamma_2$
and $\Gamma_3$ at the GUT scale, at $\lambda_{12}^E = 30 \,^{\circ} $ and
$tan(\beta)=50$. All the angles are in degrees. }

\label{tab1}

\end{table}

\begin{table}[hbtp]
    \centering

    \begin{tabular}{|l|c|c|c|c|c|}

    \hline

    $\lambda_{12}^E$& 0 & 7.5& 15 & 30 & 45  \\

    \hline

    $\theta_{12}$  & 31.72   & 31.752 & 31.846 & 32.216 & 32.8  \\

    \hline

   $\theta_{13}$ & 3.534 & 3.534 & 3.534 & 3.534 & 3.534\\

    \hline

   $\theta_{23}$ & 44.892 & 44.892 & 44.892 & 44.892 & 44.892\\

     \hline

    $\delta $& 180& 187.9& 195.789 & 211.492 & 227.039\\

    \hline
    $\Gamma_1$& 35.262 & 35.262& 35.262&35.262 &35.262 \\
    \hline
   $\Gamma_2$& 35.262 & 35.259 & 35.250 & 35.217 & 35.174 \\
    \hline
   $\Gamma_3$& 35.254 & 35.253 & 35.248 & 35.230 & 35.208\\

    \hline

    \end{tabular}

    \caption{ Values of the parameters:  $\theta_{12}$, $\theta_{13}$,$\theta_{23}$,
$\delta$ and the $\Gamma$ sum rules at the GUT scale. These values are found
in degrees at $\theta_{12}^E = 5\,^{\circ}$ and $\tan(\beta)=50$. }

    \label{tab2}

\end{table}

\subsection{More general charged lepton corrections including $\theta_{23}^E$}
In the previous subsection we saw that the sum rules
arising from Cabibbo-like charged lepton corrections are satisfied
to excellent precision at the GUT scale, for the considered numerical example.
In this section we introduce the case of
non-Cabibbo-like charged lepton corrections. To be precise we shall
consider more general charged lepton corrections given by,
\begin{equation}\label{nocabibo}
V_{e_L} \approx \left(\begin{array}{ccc} c_{\theta_{12}^E} & - s_{\theta_{12}^E}
e^{- i \lambda_{12}^E}& 0\\
s_{\theta_{12}^E} e^{ i \lambda_{12}^E} & c_{\theta_{12}^E} & 0\\
0 &0 &1
\end{array} \right)~ \left(\begin{array}{ccc} 1 &0 &0\\
0 &  c_{\theta_{23}^E} & - s_{\theta_{23}^E} e^{- i \lambda_{23}^E}\\
0 &  s_{\theta_{23}^E} e^{ i \lambda_{23}^E} & c_{\theta_{23}^E}  \end{array}
\right),
\end{equation}
where we have now allowed both $\theta_{23}^E$ and $\lambda_{23}^E$ to be non
zero. The neutrino Yukawa matrix will be transformed
to the diagonal charged lepton basis according to
\begin{equation}\label{trans}
Y_\nu \rightarrow Y_\nu' = V_{e_L} Y_\nu,
\end{equation}
but now using the non-Cabibbo-like charged lepton rotations in Eq.\ref{nocabibo}.
After performing the charged lepton rotations in Eq.\ref{trans}, values for
the mixing angles as well as the $\xi_i$ parameters given by Eq.\ref{xi1} can be
calculated at the GUT scale. Of course in the present case of non-Cabibbo-like
 charged lepton corrections the third row deviation parameters
$\xi_1$, $\xi_2$ and $\xi_3$ are all expected to be non-zero at the GUT
scale. This implies that the sum rules given by Eq.\ref{xi0} no longer apply in the case of
charged lepton corrections with non-zero $\theta_{23}^E$.
The effects of non-Cabibbo-like charged lepton corrections on the deviation
parameters $\xi_i$ is displayed in Table~\ref{tab30} using the original
neutrino Yukawa matrix as before, namely Eq.\ref{cor}, but now with a small non-zero
value of  $\theta_{23}^E=2^{\circ}$, and with different values of the new phase
$\lambda_{23}^E$.

Note that the effect of turning on the charged lepton correction $\theta_{23}^{E}$
will lead to a correction of the physical lepton mixing angle $\theta_{23}$
but not $\theta_{12}$ (to leading order) \cite{King:2005bj}.
Therefore while the sum rules $\Gamma_{1,2}$ and $\sigma_2$ are violated by
a non-zero $\theta_{23}^{E}$, the sum rules $\Gamma_{3}$ and $\sigma_1$ are both insensitive
to $\theta_{23}^{E}$.
\footnote{The insensitivity of the
sum rule $\sigma_1$ to $\theta_{23}^{E}$ is clearly seen numerically in Fig.\ref{fig:nocabidev} (b).}

\begin{table}[hbtp]

    \centering

    \begin{tabular}{|l|c|c|c|}

    \hline

    $\lambda_{23}^E ( \,^{\circ})$ & $|\xi_1|$& $|\xi_2|
$&$|\xi_3| $ \\

    \hline

   0   & 0.034 & 0.034 & 0.035 \\

    \hline

  30   & 0.027& 0.031 & 0.030\\

   \hline

    \end{tabular}

    \caption{ This table shows the values of $ |\xi_1|$,$ |\xi_2| $ and $|\xi_3| $ at
the GUT scale for case of non-Cabibbo-like charged lepton corrections with
$\theta_{12}^E = 5 \,^{\circ}$, $\lambda_{12}^E = 30 \,^{\circ}$,  $
\theta_{23}^E = 2 \,^{\circ}$ and $\tan(\beta)=50$, for different values of
the phase $\lambda_{23}^E$. }

    \label{tab30}

\end{table}

\section{Renormalization group running effects}\label{sth}
If TB neutrino mixing holds in the framework of some unified theory,
then typically we expect Cabibbo-like charged lepton corrections leading to
sum rule relations. However, as already indicated, such sum rules are only strictly
valid at the GUT scale, and will be subject to RG corrections.
In this section we now turn to a quantitative discussion of such RG corrections
to the sum rules. For definiteness we shall
assume the minimal supersymmetric standard model (MSSM), with a
SUSY breaking scale of 1 TeV, below which the SM is valid.
To study the running of the neutrino mixing angles and sum rules from the GUT
scale to the electroweak scale, the Mathematica package REAP
(Renormalization of Group Evolution of Angles and Phases) was used
\cite{Antusch:2005gp}. This package numerically solves the RGEs of the
quantities relevant for neutrino masses and mixing. It can be downloaded from
http: // www.ph.tum.de/~rge/REAP/. Mathematica 5.2 is required.

\subsection{Sum rules with Cabibbo-like charged lepton corrections}

\subsubsection{Sum rules in terms of lepton mixing angles}

In this section, we study the RG running of the sum rules
which result from Cabibbo-like charged corrections.
The neutrino Yukawa matrix is taken to be of the
form of Eq.\ref{cor} as before.
The RG change in the quantities, defined for a parameter $P$ as
$\Delta P = P_{M_Z} - P_{M_{GUT}} $,
was calculated for the lepton mixing parameters and the $\Gamma_i$ sum rules, and
is presented in Tables.\ref{tab31},\ref{tab41}. From the results we see that
the least precise sum rule $\Gamma_3$ actually is subject to the smallest RG running
since it does not involve $\theta_{23}$ which runs the most.

\begin{figure}[hbtp]
\begin{center}
\subfloat[]{\label{fig:sum}\includegraphics[height=56mm,width=81mm]
{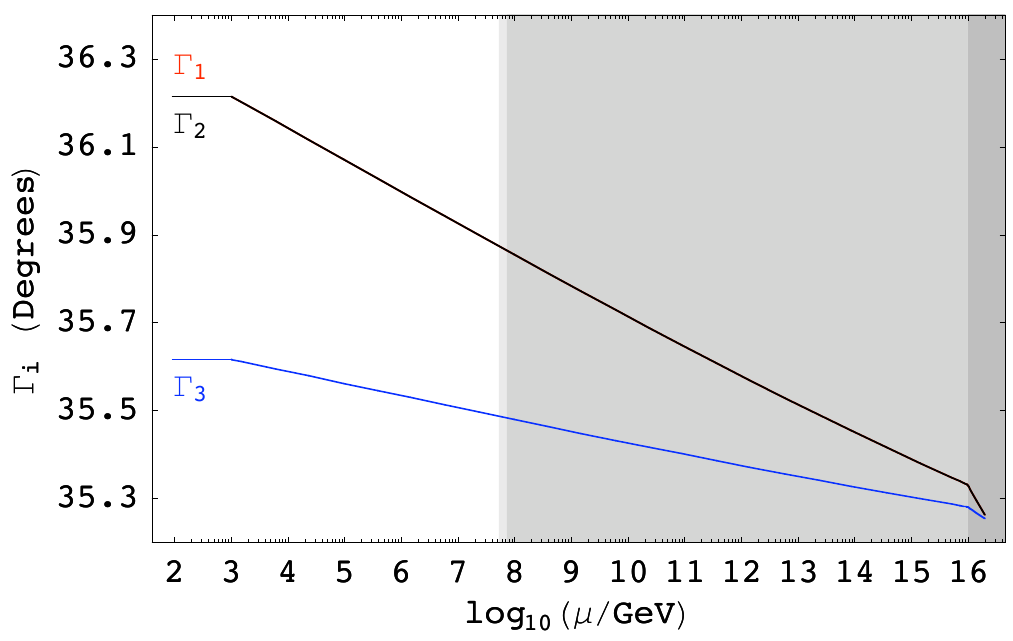}}~
\subfloat[]{\label{fig:sumh}\includegraphics[height=56mm,width=81mm]
{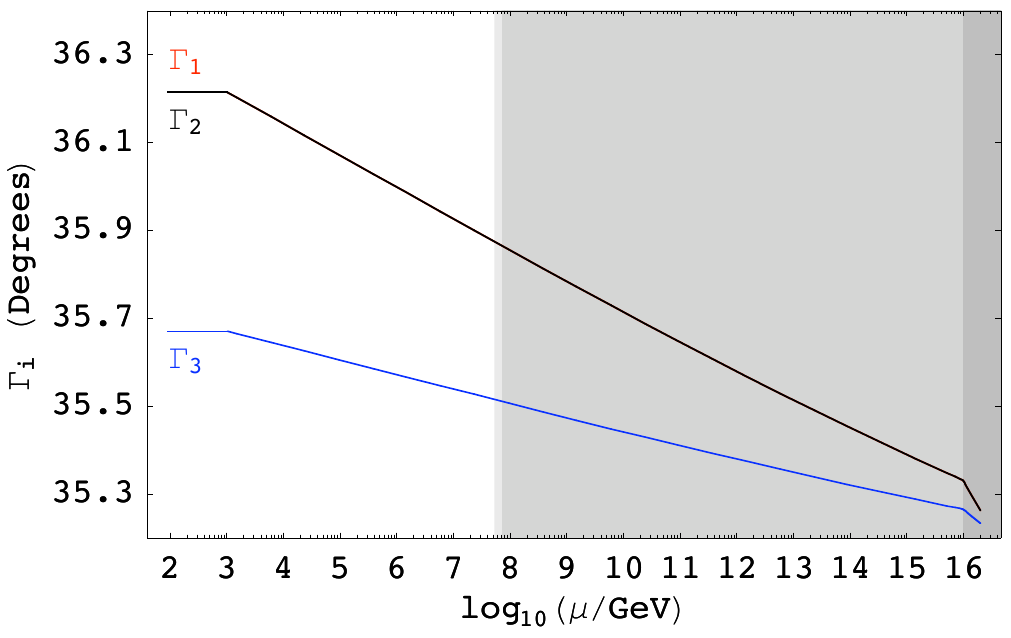}}\\
 \subfloat[]{\label{fig:sumoo}\includegraphics[height=56mm,width=81mm]
{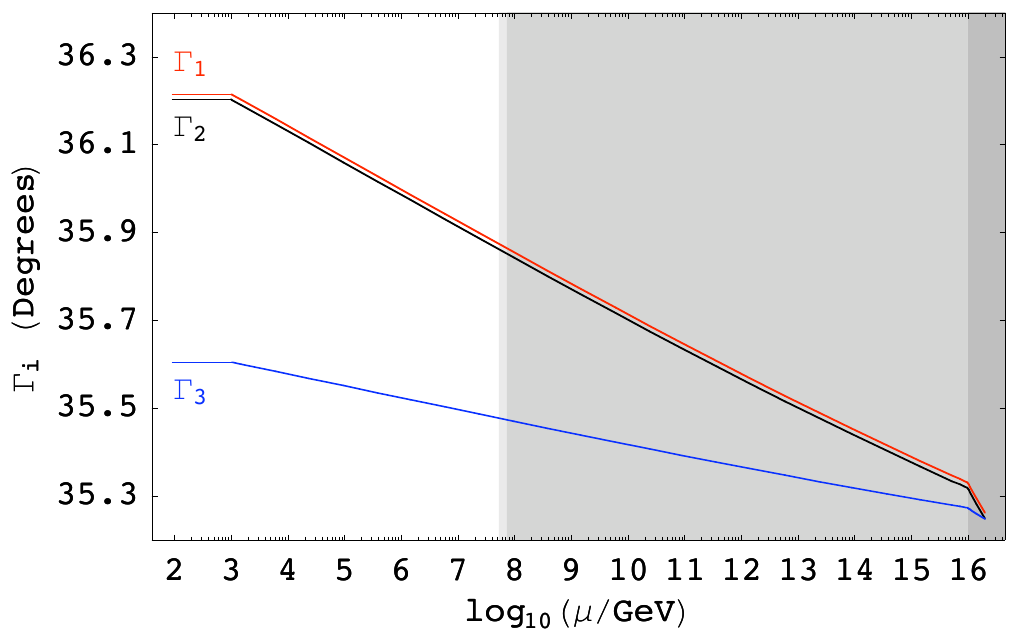}}~
 \subfloat[]{\label{fig:sumff}\includegraphics[height=56mm,width=81mm]
{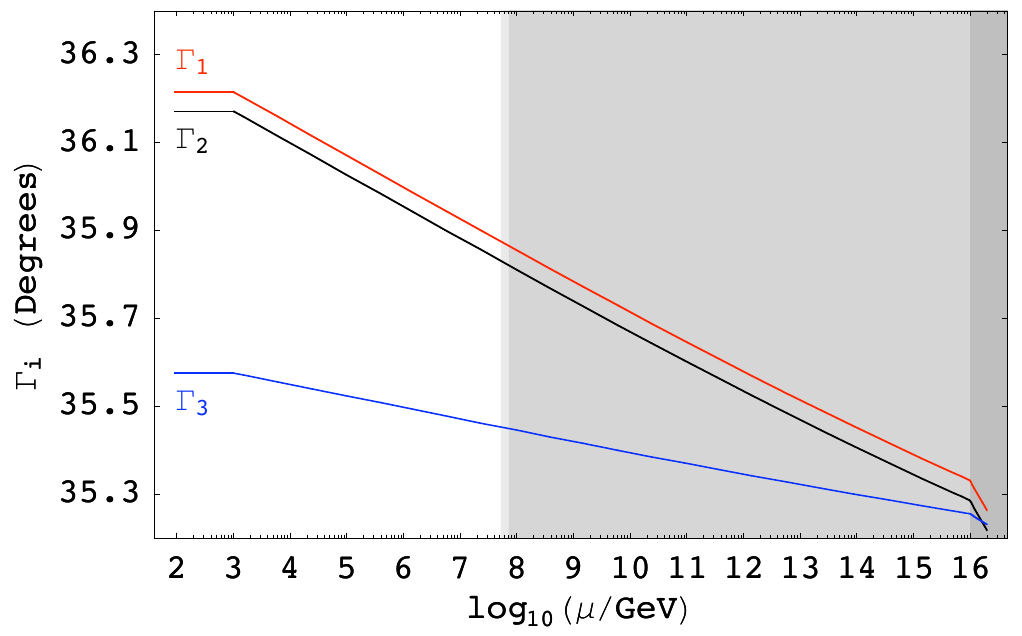}}
\end{center}
\caption{Evolution of sum rules $\Gamma_1$,$ \Gamma_2$, $
\Gamma_3$ for Cabibbo-like charged lepton corrections for large $\tan(\beta)=50$.
Panel \protect\subref{fig:sum} shows the running at $\theta_{12}^E = 5
\,^{\circ}$ and $\lambda_{12}^E = 0 \,^{\circ}$,
while panel \protect\subref{fig:sumh} shows the evolution at $\theta_{12}^E =  8 \,^{\circ}$ and
$\lambda_{12}^E = 0\,^{\circ}$. In panel \protect\subref{fig:sumoo}, the running is at $
\theta_{12}^E =5 \,^{\circ}$ and $\lambda_{12}^E= 15 \,^{\circ}$ while  panel \protect
\subref{fig:sumff} shows the evolution at  $\theta_{12}^E =5 \,^{\circ}$, $
\lambda_{12}^E= 30 \,^{\circ}$. Note how the graphs for $\Gamma_1$ and $
\Gamma_2$ completely overlap in \protect\subref{fig:sum} and \protect
\subref{fig:sumh}.}
\label{fig:sum1}
\end{figure}

\begin{figure}[hbtp]
\begin{center}
 \subfloat[  ]{\label{fig:tanb}  \includegraphics[height=56mm,width=80mm]
{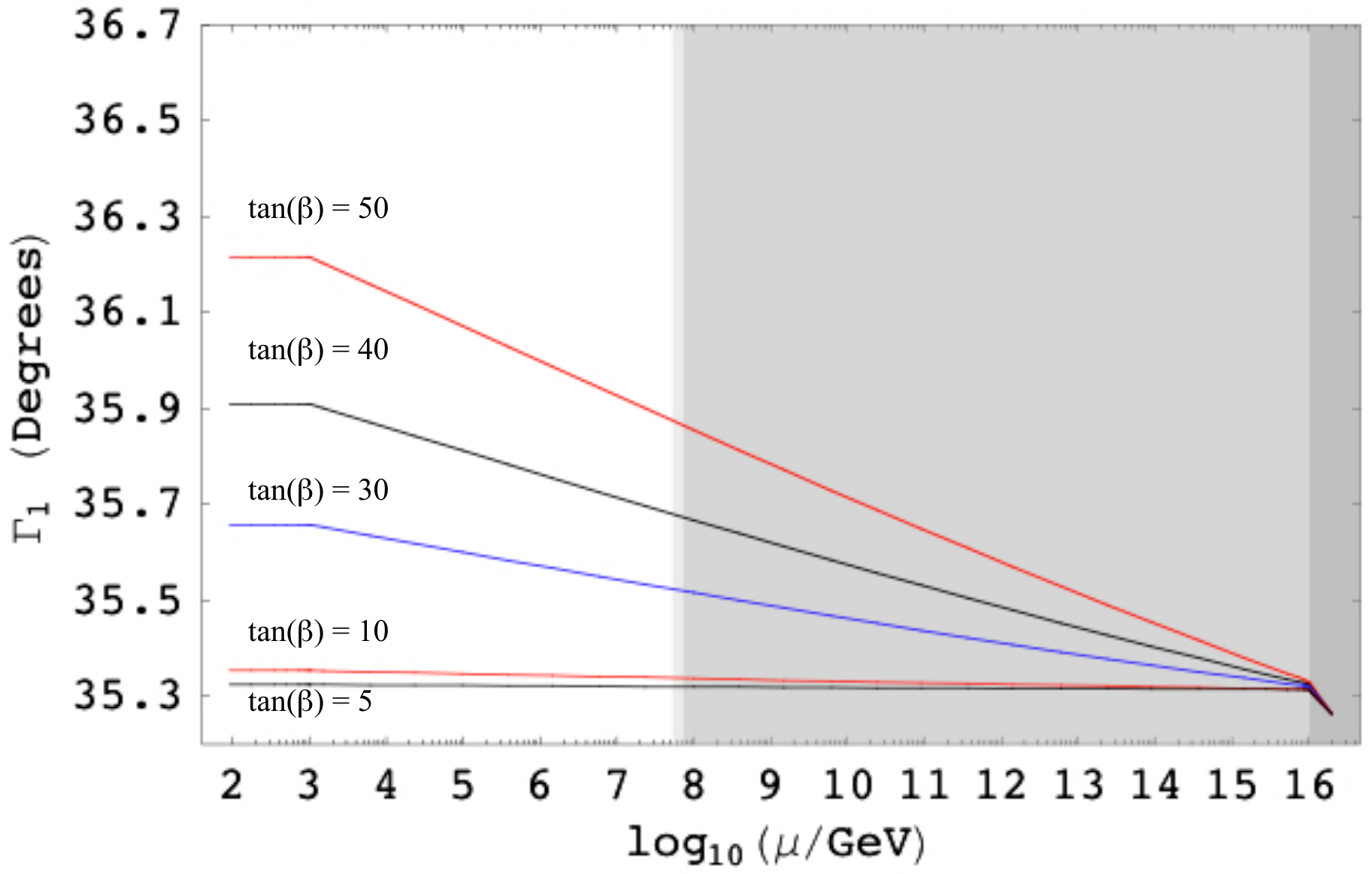}}~
\subfloat[ ]{\label{fig:tanb1}  \includegraphics[height=56mm,width=80mm]
{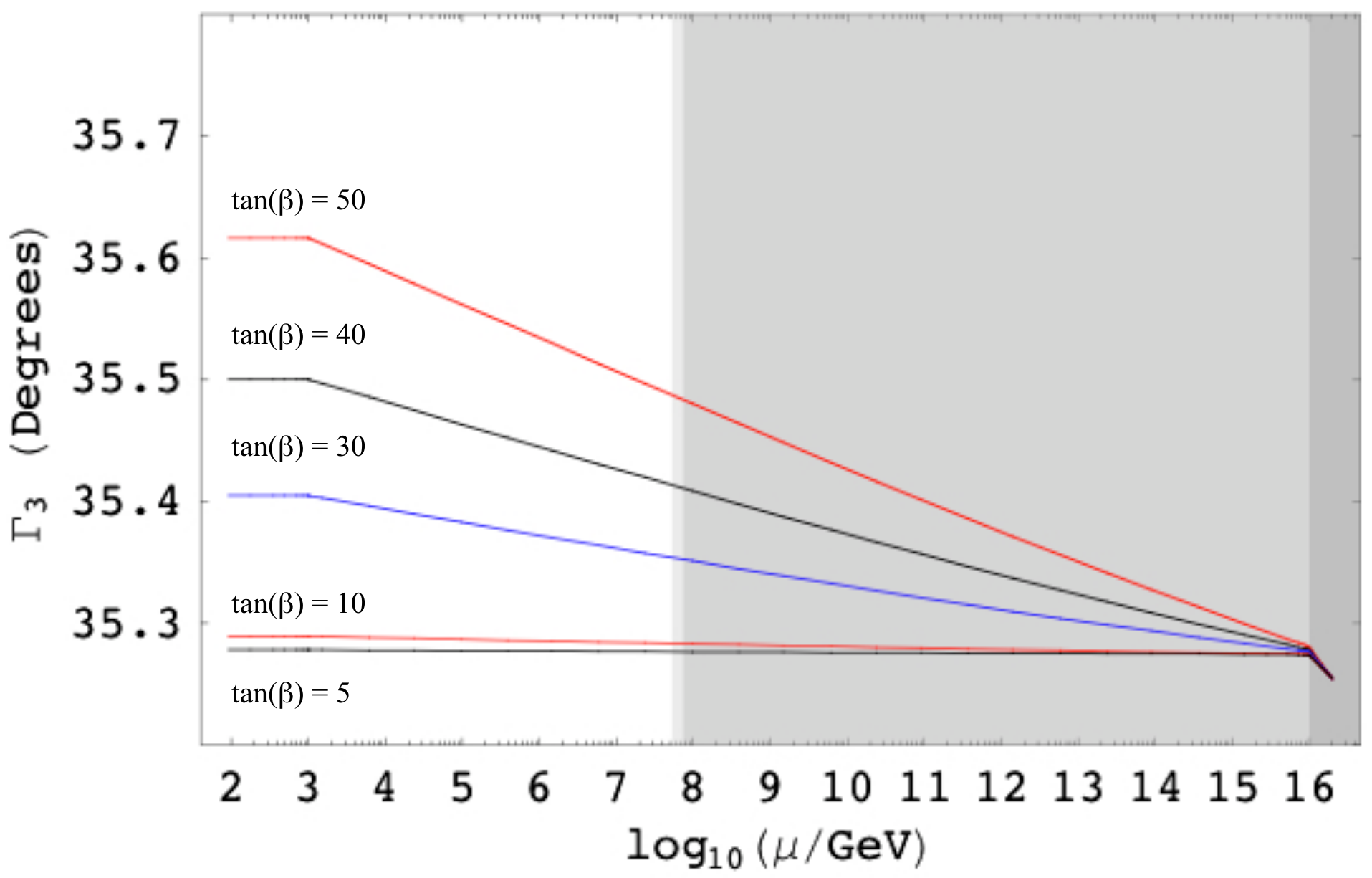}}
\end{center}
\caption{Evolution of sum rules $\Gamma_1$, $
\Gamma_3$ for Cabibbo-like charged lepton corrections for various $\tan(\beta)$.
Running of $\Gamma_1$  is shown in panel \protect\subref{fig:tanb} and $\Gamma_3$
is shown in panel \protect\subref{fig:tanb1} at  $\theta_{12}^E =5
\,^{\circ}$, $\lambda_{12}^E= 0 \,^{\circ}$ and different values of $\tan(\beta)$.
Note the expanded (and different) vertical scales used in these two figures;
in all cases the corrections are less than one degree.}
\label{fig:tb}
\end{figure}

The RG running of $\Gamma_i$ is displayed in
Fig.\ref{fig:sum1} for $\tan(\beta)=50$.
The RG evolution of $\Gamma_1$ and $\Gamma_3$ was also plotted at
 different values of $\tan(\beta)$ as shown in Fig.\ref{fig:tb}.

\begin{table}[hbtp]

    \centering

    \begin{tabular}{|l|c|c|c|c|c|}

    \hline

    $\theta_{12}^E$& 0 & 1& 3 & 5 & 8  \\

    \hline

    $\Delta \theta_{12}$  & +0.391   & +0.402 & + 0.423  & + 0.444 & + 0.473 \\

    \hline

   $\Delta \theta_{13}$ & + 0.151 & - 0.116 & - 0.095 & - 0.071 & - 0.033\\

    \hline

   $\Delta \theta_{23}$ & + 1 & + 1.001 & + 1.004 & + 1.008 & + 1.013\\

     \hline

    $\Delta \delta $& 0 & + 7.453 & + 2.126 & + 1.181 & + 0.62\\

    \hline
    $\Delta \Gamma_1$& + 0.953 & + 0.953 & + 0.953& + 0.953 & + 0.953 \\
    \hline
   $\Delta \Gamma_2$& + 0.953 & + 0.953& + 0.953 & + 0.954 & + 0.958 \\
    \hline
   $\Delta \Gamma_3$& + 0.237 & + 0.259 & + 0.301 & + 0.345 & + 0.412\\

    \hline

    \end{tabular}

    \caption{ RG changes of the mixing parameters and  sum rules $\Gamma_1$,
$
\Gamma_2$ and $\Gamma_3$ at $\lambda_{12}^E = 30 \,^{\circ}$ and
$tan(\beta)=50$. All values are in degrees}

    \label{tab31}

\end{table}

 \begin{table}[hbtp]

    \centering

    \begin{tabular}{|l|c|c|c|c|c|}

    \hline

    $\lambda_{12}^E$& 0 & 7.5& 15 & 30 & 45  \\

    \hline

    $\Delta \theta_{12}$  & + 0.454  & + 0.453 & + 0.452 & + 0.444 & + 0.432  \\

    \hline

   $\Delta \theta_{13}$ & - 0.092 & - 0.091 & - 0.087  & - 0.071  & - 0.046\\

    \hline

   $\Delta \theta_{23}$ & + 1.009 & + 1.009 & + 1.009 & + 1.008 & + 1.006\\

     \hline

    $\Delta \delta $& 0 & + 0.31 & + 0.613 & + 1.181 & + 1.663\\

    \hline
    $\Delta \Gamma_1$& + 0.953 & + 0.953& + 0.953& + 0.953 & + 0.953 \\
    \hline
   $\Delta \Gamma_2$& + 0.953 & + 0.953 & + 0.953 & + 0.954 & + 0.956 \\
    \hline
   $\Delta \Gamma_3$& + 0.362 & + 0.36 & + 0.357 & + 0.345 & + 0.326 \\

    \hline

    \end{tabular}

    \caption{ RG changes of the neutrino mixing angles, the Dirac phase $\delta$
and the sum rules $\Gamma_1$, $\Gamma_2$ and $\Gamma_3$ at $
\theta_{12}^E = 5\,^{\circ}$ and $tan(\beta)=50$. All values are in degrees.}

    \label{tab41}

\end{table}

\subsubsection{Sum rules in terms of TB deviation parameters}\label{s2}

\begin{figure}[hbtp]
\begin{center}
 \subfloat[ ]
{ \label{fig:dev0} \includegraphics[height=56mm,width=81mm]
{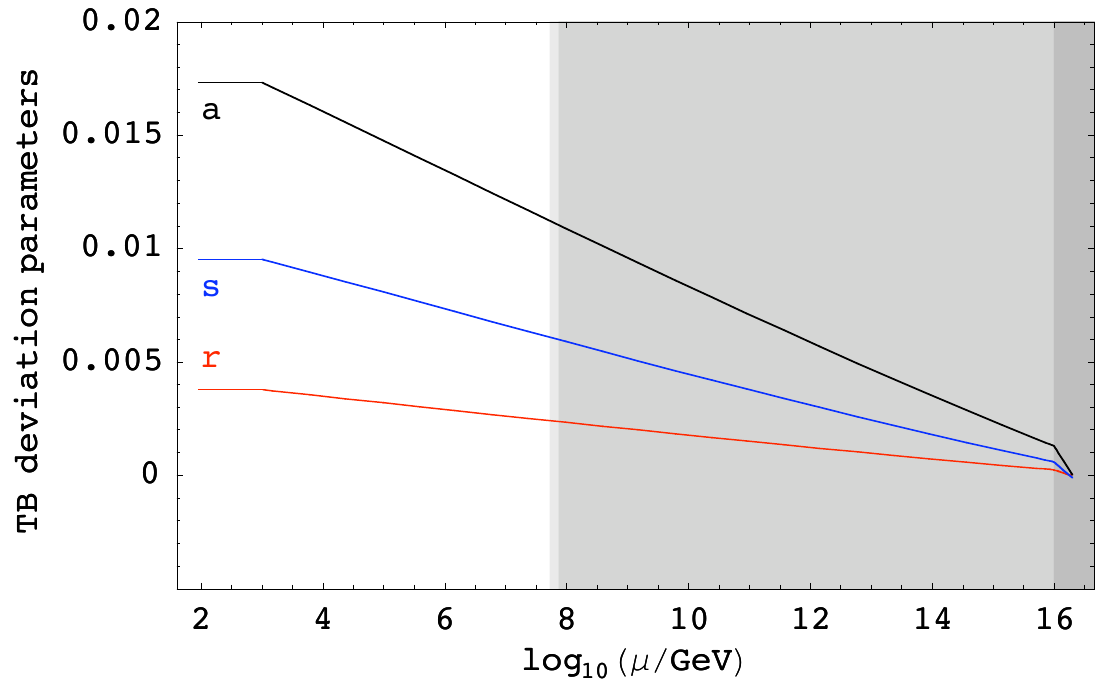}}~~~~~~~
\subfloat[ ]
{ \label{fig:dev1} \includegraphics[height=56mm,width=81mm]{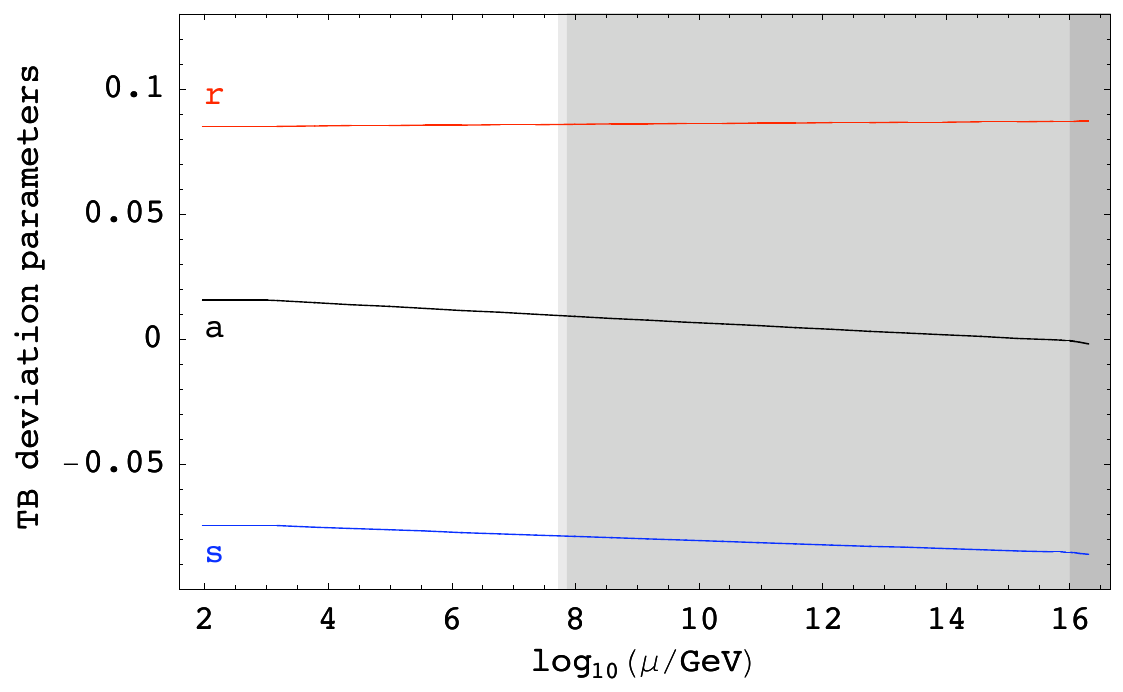}}
\end{center}
\caption{Evolution of the deviation parameters $r$, $s$, $a$ from the GUT scale
to the electroweak scale without and with charged lepton corrections
for large $\tan(\beta)=50$.
Panel \protect\subref{fig:dev0} shows the
running of the deviation parameters without any charged lepton corrections,
$\theta_{12}^E =0\,^{\circ}$, $
\lambda_{12}^E= 0 \,^{\circ}$.
Panel \protect\subref{fig:dev1} shows the
running of the deviation parameters in the presence of
Cabibbo-like charged lepton corrections,
$\theta_{12}^E =5 \,^{\circ}$, $
\lambda_{12}^E= 15 \,^{\circ}$.}
\label{fig:devi}
\end{figure}

\begin{figure}[hbtp]
\begin{center}
 \subfloat[  ]{\label{fig:sdo} \includegraphics[height=56mm,width=81mm]
{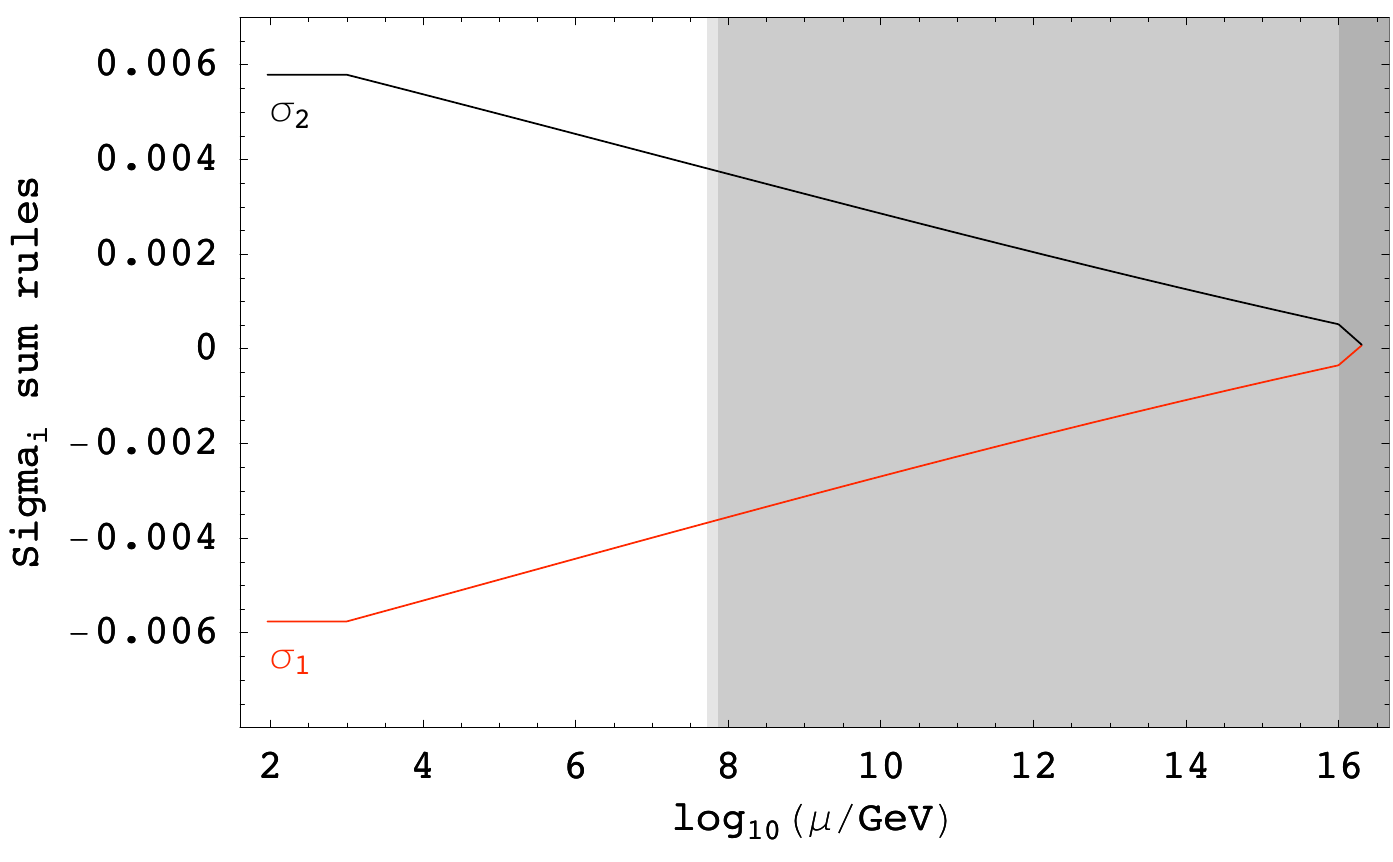}}~
\subfloat[ ]{\label{fig:sdt} \includegraphics[height=56mm,width=81mm]
{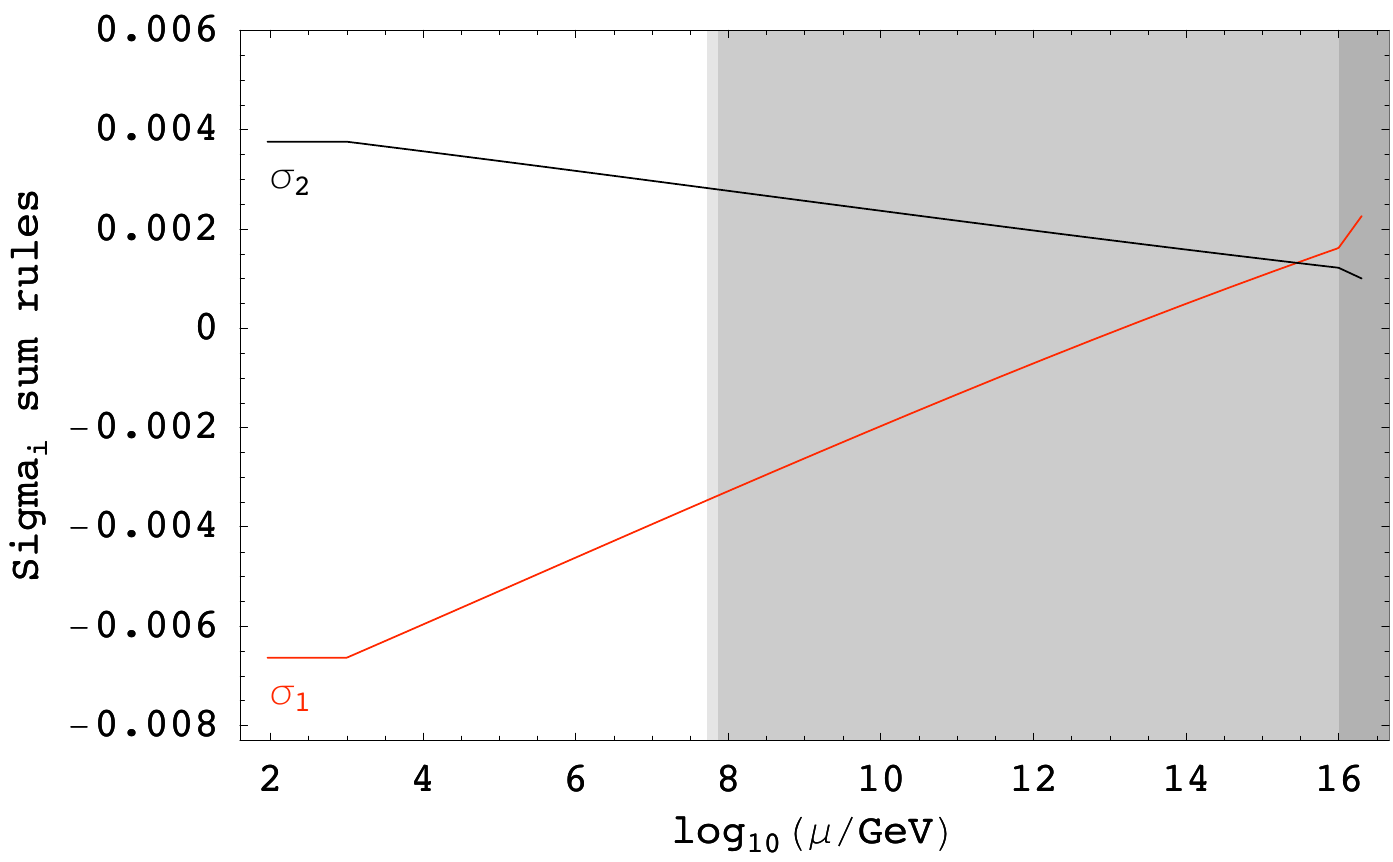}}
\end{center}
\caption{Evolution of the sum rules $\sigma_1$ and  $\sigma_2$  from
the GUT scale to the electroweak scale without and with charged lepton corrections
for large $\tan(\beta)=50$. Panel \protect\subref{fig:sdo}
 shows the running without any charged lepton corrections,
 $\theta_{12}^E = 0\,^{\circ}$ and $\lambda_{12}^E = 0\,^{\circ}$.
 In panel \protect\subref{fig:sdt} the evolution is
in the presence of
Cabibbo-like charged lepton corrections,
 $\theta_{12}^E = 5\,^{\circ}$ and $
\lambda_{12}^E = 30\,^{\circ}$}
\label{fig:smdev}
\end{figure}

In this subsection, for completeness we study the evolution of the TB deviation parameters
defined in Eq.\ref{de}. Their RG evolution, for different values of $\theta_{12}^E$, is
shown in Fig.\ref{fig:devi}. In Fig.\ref{fig:smdev}
we display the evolution of the sum rules given by Eqs.\ref{sd1}, \ref{sd2}.
From Fig.\ref{fig:smdev} it is seen that both $\sigma_1$, $\sigma_2$ are precisely equal to
zero at the GUT scale for $\theta_{12}^E = 0$ but differ by a tiny amount for $ \theta_{12}^E,
\lambda_{12}^E \not= 0$. In this numerical example it is apparent that
the sum rule $\sigma_2$ is slightly more stable than the
original sum rule $\sigma_1$, although there is not much more stability.
This is a manifestation of the fact that $\sigma_2$ does not take
into account the running of $r$, which introduces an effect coming from
the Majorana phases which we have assumed to be zero in this example.
Later on we shall discuss a numerical example with non-zero Majorana
phases where the enhanced stability of $\sigma_2$ will be more pronounced.

\subsection{Sum rules with more general charged lepton corrections including $\theta_{23}^E $}

\begin{figure}[hbtp]
\begin{center}
 \subfloat[  ]{\label{fig:xim1} \includegraphics[height=56mm,width=81mm]
{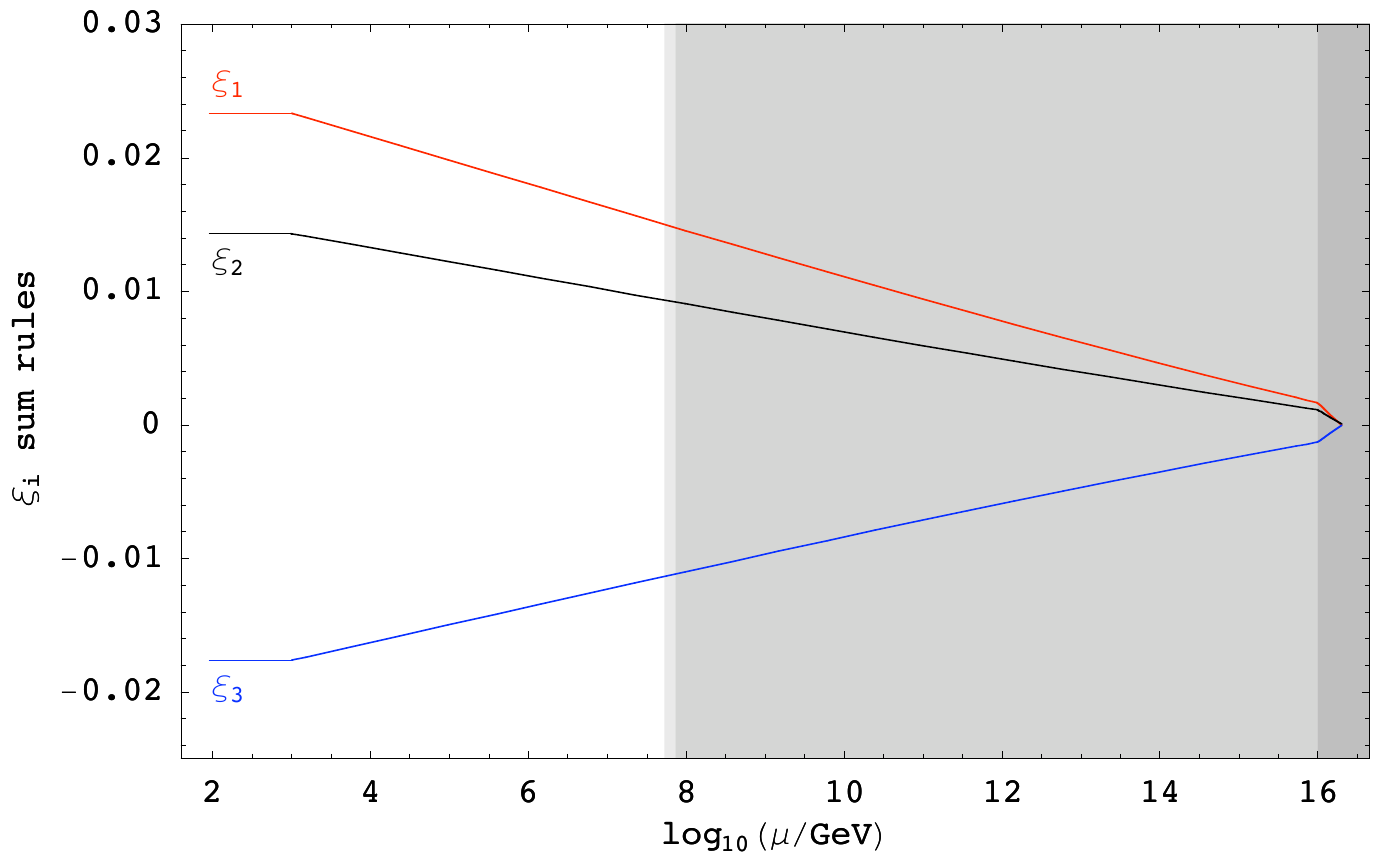}}~~~~~~~
\subfloat[ ]{\label{fig:xim2} \includegraphics[height=56mm,width=81mm]
{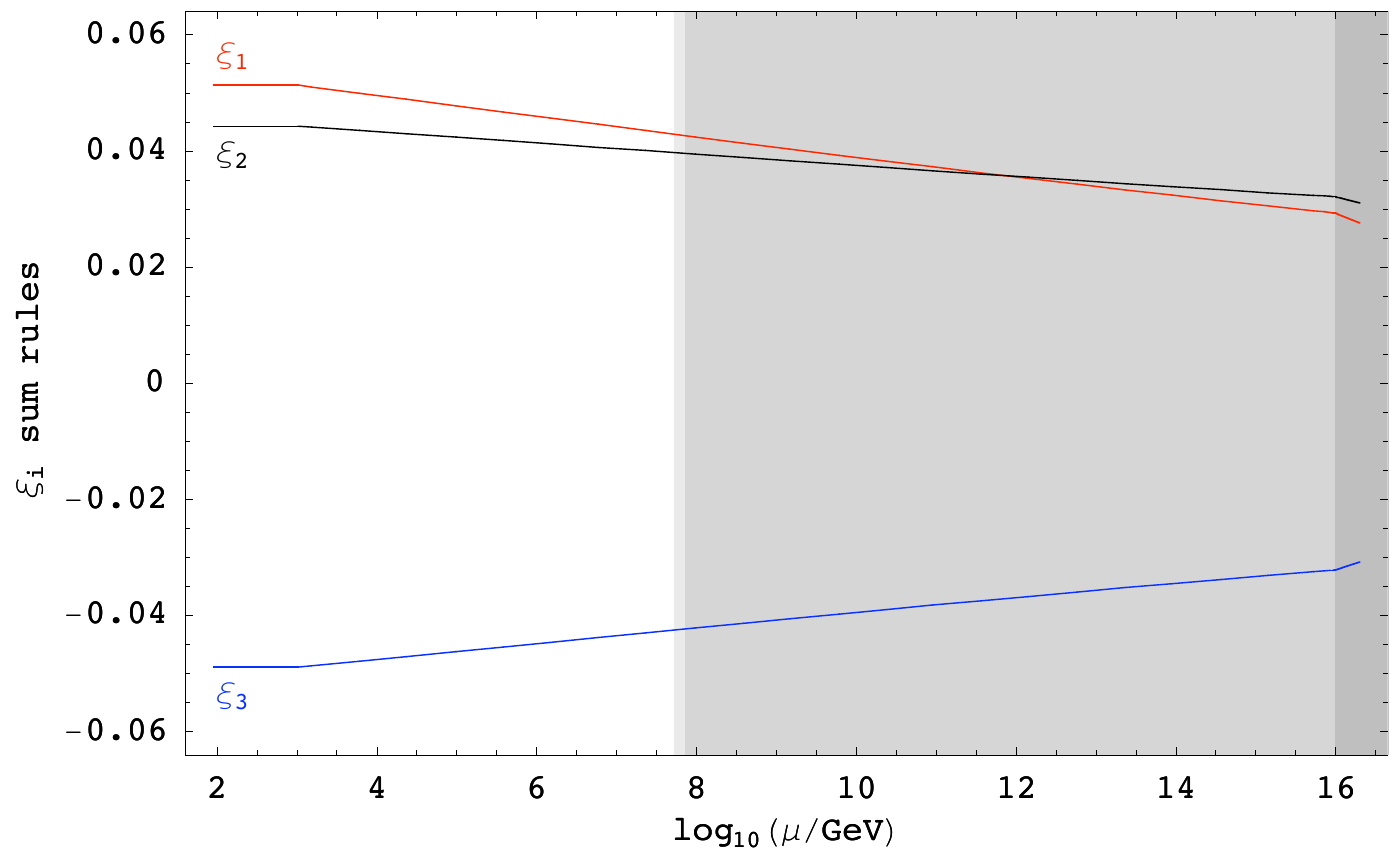}}
\end{center}
\caption{The evolution of the third row deviation parameters
$\xi_1$, $\xi_2$ and $\xi_3$ from the GUT scale to the
electroweak scale at  $\tan(\beta)=50$
for Cabibbo-like and more general charged lepton corrections.
Panel \protect\subref{fig:xim1} shows the result with
Cabibbo-like charged lepton corrections with $\theta_{23}^E =
0\,^{\circ}$, $\lambda_{23}^E = 0\,^{\circ}$, $\theta_{12}^E = 5\,^{\circ}$ and $
\lambda_{12}^E = 30\,^{\circ}$. Panel \protect\subref{fig:xim2}
is for the case of more general charged lepton corrections with
$\theta_{23}^E = 2\,^{\circ}$ , $
\lambda_{23}^E = 30\,^{\circ}$, $\theta_{12}^E = 5\,^{\circ}$ and $
\lambda_{12}^E
= 30\,^{\circ}$.} \label{fig:xidevm2}
\end{figure}

\begin{figure}[hbtp]
\begin{center}
 \subfloat[  ]{\label{fig:ns1} \includegraphics[height=56mm,width=81mm]
{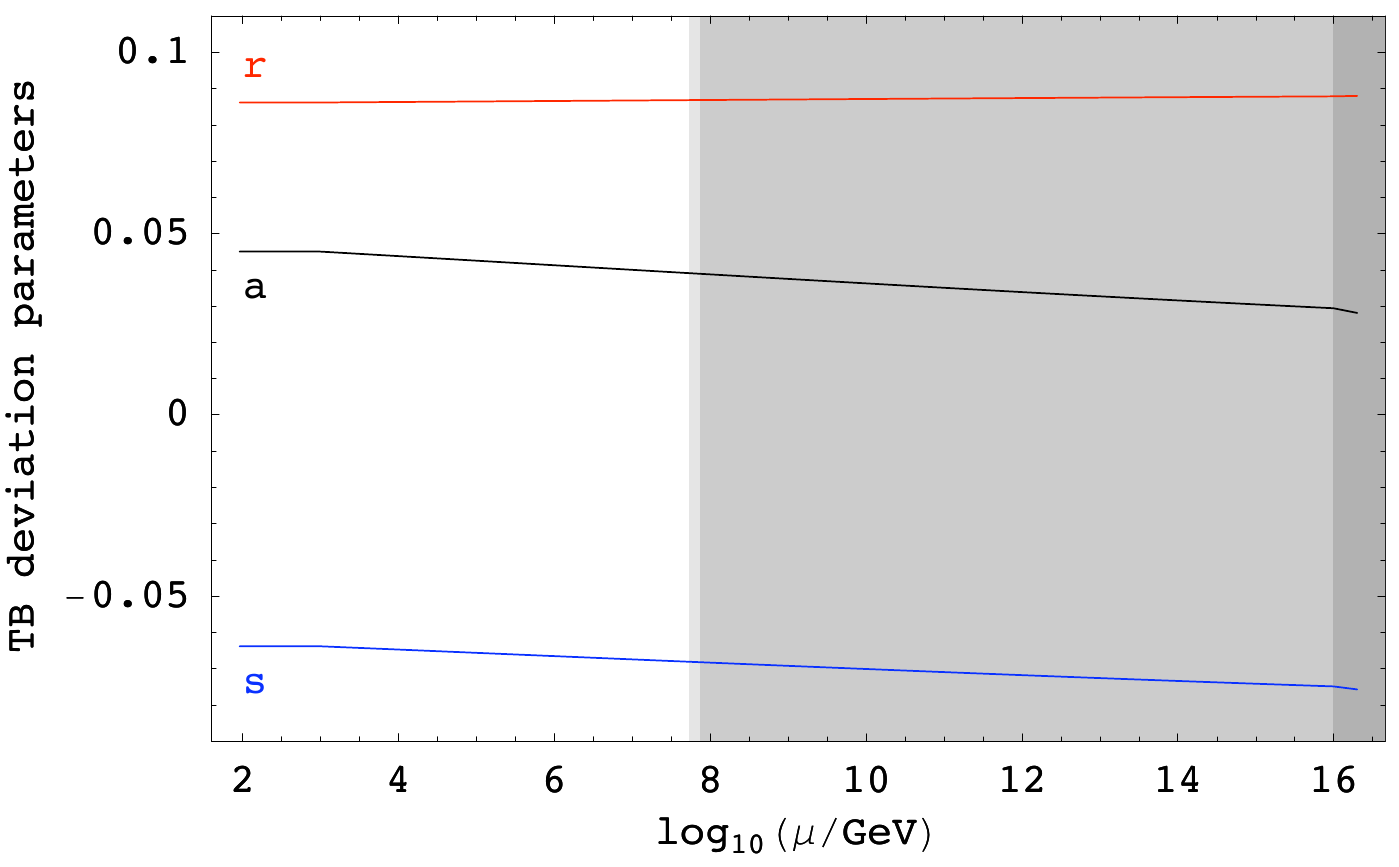}}~
\subfloat[ ]{\label{fig:ns2} \includegraphics[height=56mm,width=81mm]
{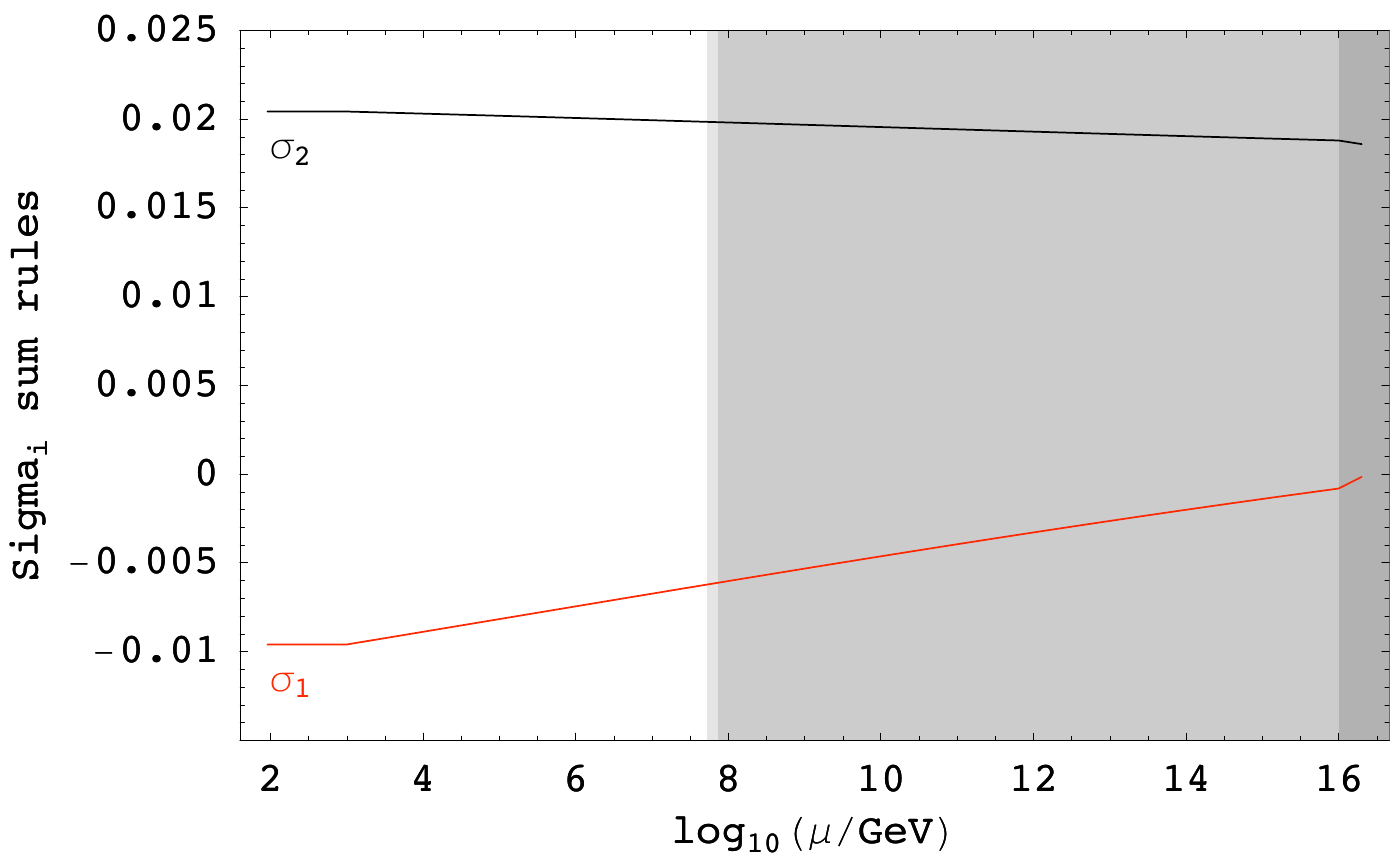}}
\end{center}
\caption{Running of the TB deviation parameters and their related sum rules
from the GUT scale to the electroweak scale
in the presence of more general charged lepton corrections,
$\theta_{12}^E = 5 \,^{\circ}$, $\lambda_{12}^E = 30 \,^{\circ}$, $\lambda_{23}^E =
30 \,^{\circ}$, $\theta_{23}^E = 2 \,^{\circ}$ and $\tan(\beta)=50$.
Panel \protect\subref{fig:ns1} shows the evolution of the TB deviation parameters.
Panel \protect\subref{fig:ns2} displays the sum rules
$\sigma_1$ and $\sigma_2$. Note that $\sigma_1 = 0$ at the GUT scale
even in the presence of the more general charged lepton corrections.}
\label{fig:nocabidev}
\end{figure}

Finally in this subsection
we study the evolution of the $\xi_i$ parameters for the case of
charged lepton corrections of the more general form in Eq.\ref{nocabibo}.
In Fig.\ref{fig:xidevm2} \protect \subref{fig:xim1} we
show the RG running of the parameters $\xi_1$, $\xi_2$ and $\xi_3$, given in
terms of the mixing angles in Eq.~\ref{xi2}, for the case of
Cabibbo-like charged lepton corrections.
As expected, for Cabibbo-like charged lepton corrections,
these parameters are exactly zero
at the GUT scale for all values of $\theta_{12}^E$ and $\lambda_{12}^E$, but then
diverge from zero due to the RG corrections.
In Fig. \ref{fig:xidevm2}  \protect\subref{fig:xim2} we now switch on
the non-Cabibbo-like charged lepton corrections by a small amount
corresponding to $\theta_{23}^E = 2\,^{\circ}$. In this case we see that
the parameters $\xi_1$, $\xi_2$ and $ \xi_3$
are all non zero at the GUT scale and deviate even more at low energies
due to RG running.

In Fig.\ref{fig:nocabidev} we show the running of the TB deviation parameters and the
sum rules $\sigma_1$ and $\sigma_2$ for the
non-Cabibbo-like case with $\theta_{23}^E = 2\,^{\circ}$.
It is clear from panel (b) that the $\sigma_1$ sum rule is still valid at the
GUT scale even for a non-zero $\theta_{23}$, as remarked earlier.

\newpage

\subsection{Sum rules with non-zero Majorana phases}\label{phnot0}

So far we have presented results for a particular example with zero Majorana phases.
In this section, we present the running of the $\sigma_i$ sum rules and the TB deviation
parameters where the neutrino Yukawa matrix is taken to be similar to Eq.\ref{cor} with
the same values for $b$, $e$ and $c_3$ but with non- zero Majorana phases:
\begin{equation}
\label{cor2}
Y_{LR}^\nu =\left(\begin{array}{ccc} 0& 0.97282 b e^{i \delta_2} & 0.001\\
e e^{i \delta_1} & b e^{i \delta_2} & 0\\
- 1.012 e e^{i \delta_1} & b e^{i \delta_2} & c_3\end{array} \right)
\end{equation}
where we shall take the phases to be $\delta_1 = 120^o$
and $ \delta_2 = 60^o$. The right-handed Majorana mass matrix is as before.
The numerical value of the Yukawa couplings
has been changed slightly to compensate for the
non-zero phases in order to once again yield exact tri-bimaximal
neutrino mixing at the GUT scale.

\begin{figure}[hbtp]
\begin{center}
 \subfloat[  ]{\label{fig:news} \includegraphics[height=56mm,width=81mm]
{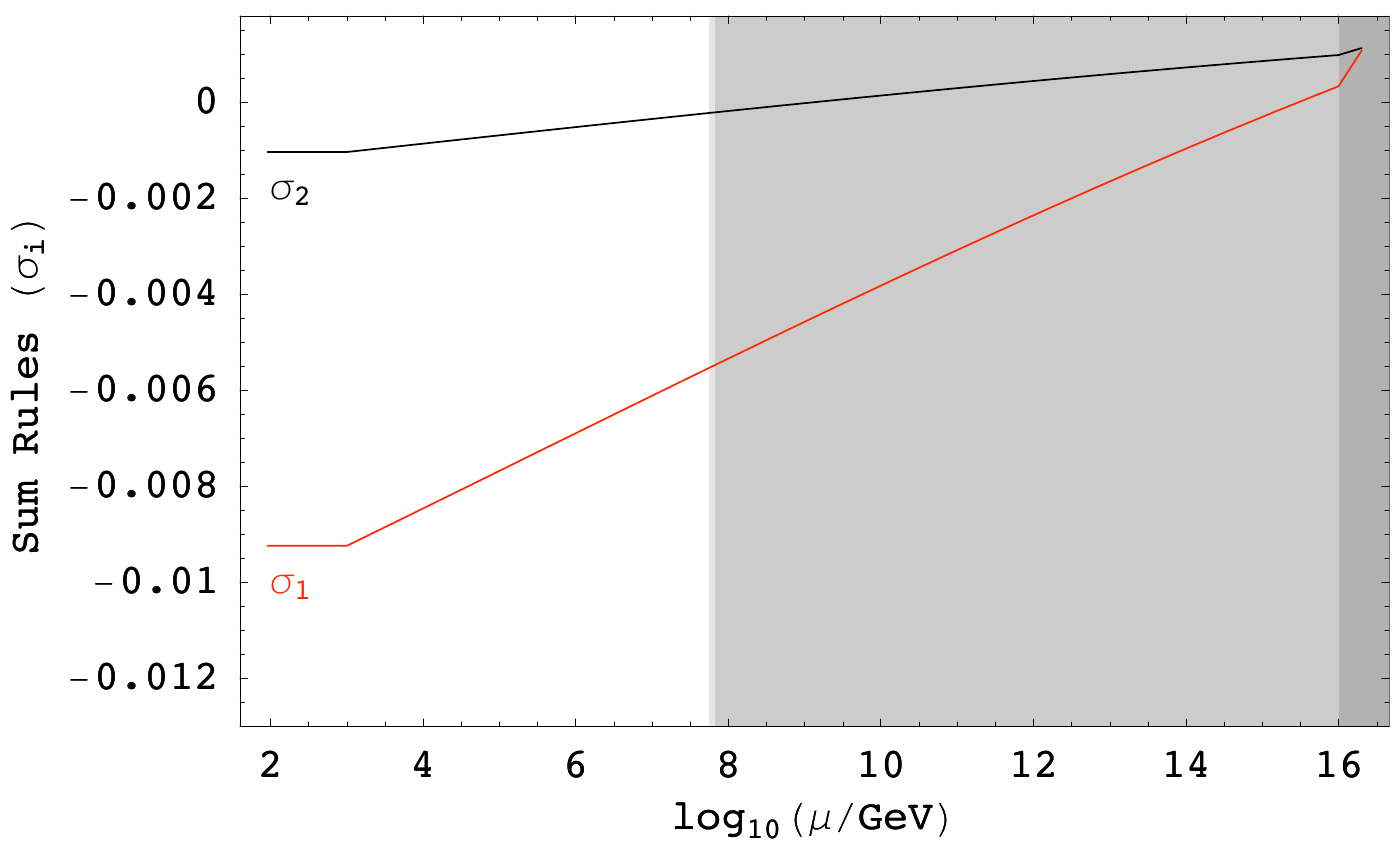}}~
\subfloat[ ]{\label{fig:newdev} \includegraphics[height=56mm,width=81mm]
{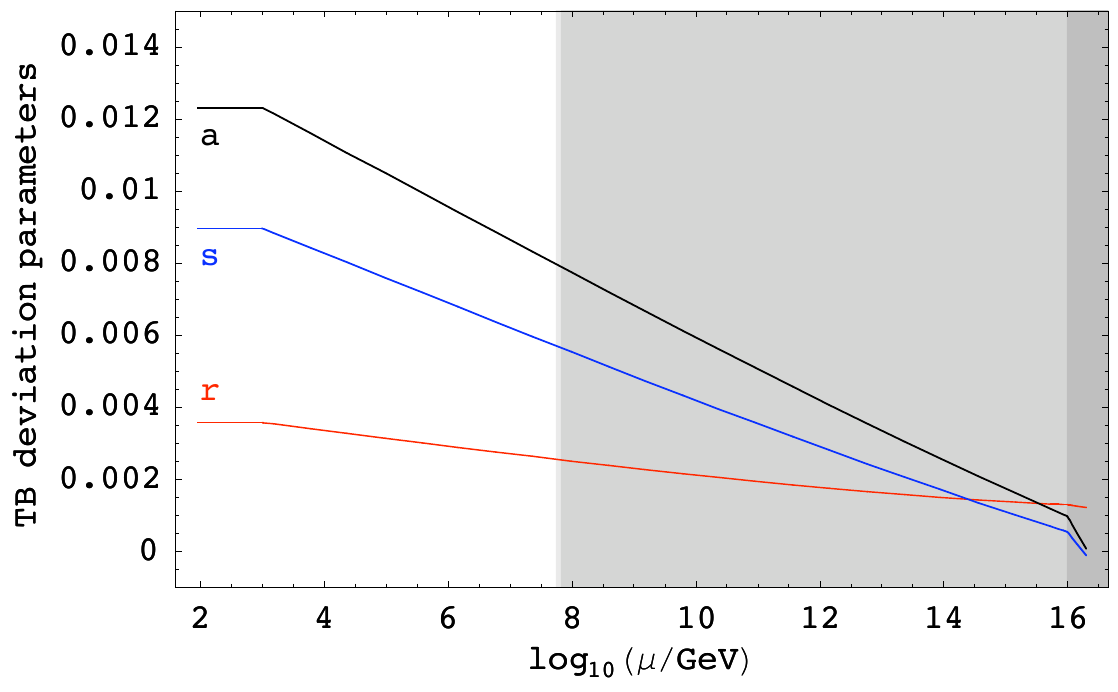}}
\end{center}
\caption{Sum rules with Majorana phases.
The running of $\sigma_i$ sum rules and TB deviation parameters are for
$\theta_{12}^E = 0 \,^{\circ}$, $\lambda_{12}^E = 0 \,^{\circ}$, $\tan(\beta)=50$,
$\delta_1 = 120^o$ and
$\delta_2 = 60^o$. }
\label{fig:sumndev}
\end{figure}

In Fig. \ref{fig:sumndev} we show results for the running of the sum rules
$\sigma_i$ and for the deviation parameters $r,s,a$ for the above example with
non-zero Majorana phases. In this example
the $\sigma_2$ sum rule is much more stable than $\sigma_1$
as clearly shown in Fig.\ref{fig:sumndev}  \protect\subref{fig:news}.
This shows that the question of the stability of the sum rule $\sigma_2$
is dependent on the choice of Majorana phases via the running of $r$.
In particular with this choice of Majorana phases the
deviation parameters $s$, $a$ and $r$ all run less as shown in Fig.\ref{fig:sumndev}
\protect\subref{fig:newdev}, compared to the previous case with zero phases (Fig.\ref{fig:devi}
\protect\subref{fig:dev0} ).

The $\Gamma_i$ and $\xi_i$ sum rules also change with the Majorana phases turned on but not
as much
as $\sigma_i$ sum rules. For instance, at $\theta_{12}^E= 5^o$ and $\lambda_{12}^E = 0^o$,
we find that $\Gamma_1$ and $\Gamma_2$ get smaller by 0.05 degrees at the $M_Z$ scale
compared to the case where the phases are zero. $\Gamma_3$ on the other hand gets larger
by about 0.1 degrees. At $\theta_{12}^E= 5^o$ and $\lambda_{12}^E = 30^o$, $\xi_1$ and
$\xi_2$ get smaller by about 0.001 to 0.003 compared to the zero phases case whereas
$\xi_3$ gets larger by 0.006.

\clearpage
\section{Model dependence of the results: heavy sequential dominance}\label{hsd}

So far all the numerical results have been based on a particular example inspired
by the models of \cite{King:2005bj,deMedeirosVarzielas:2005ax}, namely the
case where the GUT scale neutrino Yukawa matrix has the form in Eq.\ref{cor},
or the closely related form in Eq.\ref{cor2} with non-zero Majorana phases.
In these examples the dominant contribution to atmospheric neutrino mass
is coming from the lightest right-handed neutrino via the see-saw mechanism,
a situation known as light sequential dominance (LSD) \cite{King:1998jw}. In order to test the
generality of the results in this section we consider a quite different example
in which the dominant contribution to the
atmospheric neutrino mass
is coming from the heaviest right-handed neutrino via the see-saw mechanism,
a situation known as heavy sequential dominance (HSD) \cite{King:1998jw}.
This example is chosen since it the most qualitatively different to the
example of LSD considered previously, yet despite this we shall see that the
numerical results for the corrections to TB mixing are qualitatively similar to
those encountered previously. This gives us some confidence that our results
and conclusions are not restricted to the particular numerical example
studied but are in fact applicable to a large class of see-saw models
based on hierarchical neutrino masses.

In the HSD example considered here the right handed neutrino Majorana matrix as well as the
neutrino Yukawa matrix are given by the following equations:
\[
M_{RR}=\left( \begin{array}{ccc} 3.991 \times 10^{-6} & 0 & 0 \\
0 & 5.800 \times 10^{-4} & 0 \\
0 & 0 & 5.021 \end{array} \right) M_3,
\]
where $M_3 = 10^{14} GeV$.
Ignoring RGE corrections to begin with,
we find that precise tri-bimaximal neutrino mixing
at the GUT scale ($\theta^{\nu}_{12}= 35.26  \,^{\circ}$, $
\theta^{\nu}_{23} =45.00  \,^{\circ}$, $\theta^{\nu}_{13} =0.00  \,^{\circ}$)
can be achieved with the Yukawa matrix:
\begin{equation}\label{cor3}
Y_{LR}^\nu =\left(\begin{array}{ccc} 1.001 \times 10^{-7}  & 1.0036 ~b & 0\\
0 & b  & -1.0013~ e \\
 2.992 \times 10^{-5} & b  & e  \end{array} \right)
\end{equation}
where $b~= 2.401 \times 10^{-3}$, $e~= 0.677$.
These parameters also lead to
the following values for the neutrino
masses: $\Delta m^2_{atm} = 2.47 \times 10^{-3} eV^2$ and $\Delta m^2_{sol} =
7.53 \times 10^{-5} eV^2$ which are well within the allowed experimental ranges.

Note that in the case of HSD the Yukawa
couplings present in the neutrino Yukawa matrix are larger than the previous case
especially $e$ which we take to be 0.677 compared to $2.125 \times 10^{-4}$ in
the previous example. Furthermore there are similarly two large Yukawa couplings
in the third column of the Yukawa matrix. Moreover the heaviest RH neutrino
associated with these large Yukawa couplings has a mass
well below the GUT scale leading larger threshold corrections coming from it.

\begin{figure}[hbtp]
\begin{center}
 \subfloat[  ]{\label{fig:nm1} \includegraphics[height=56mm,width=81mm]
{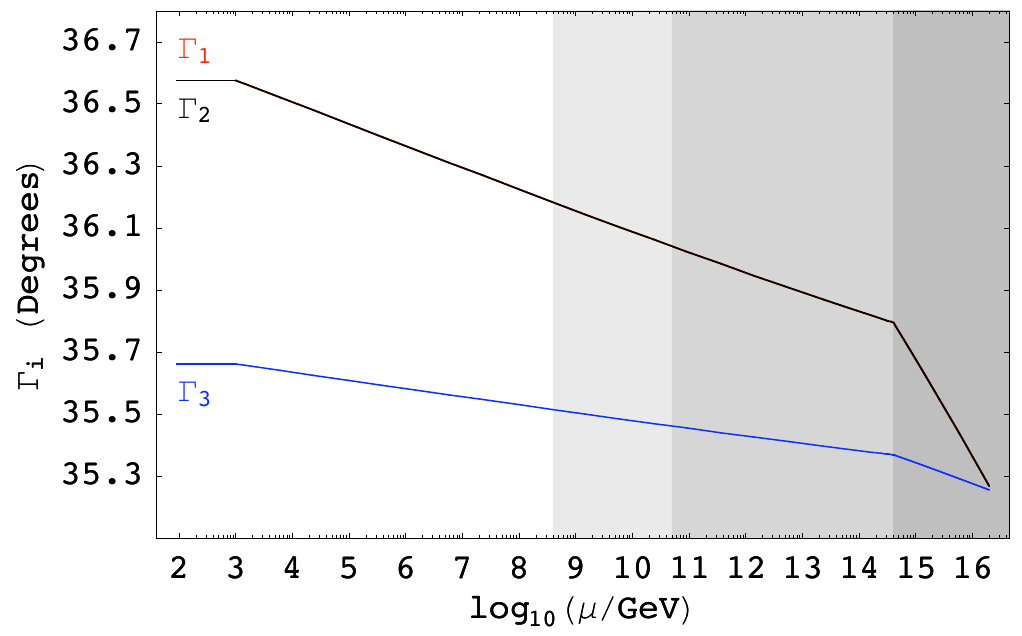}}

\subfloat[ ]{\label{fig:nm2} \includegraphics[height=56mm,width=81mm]
{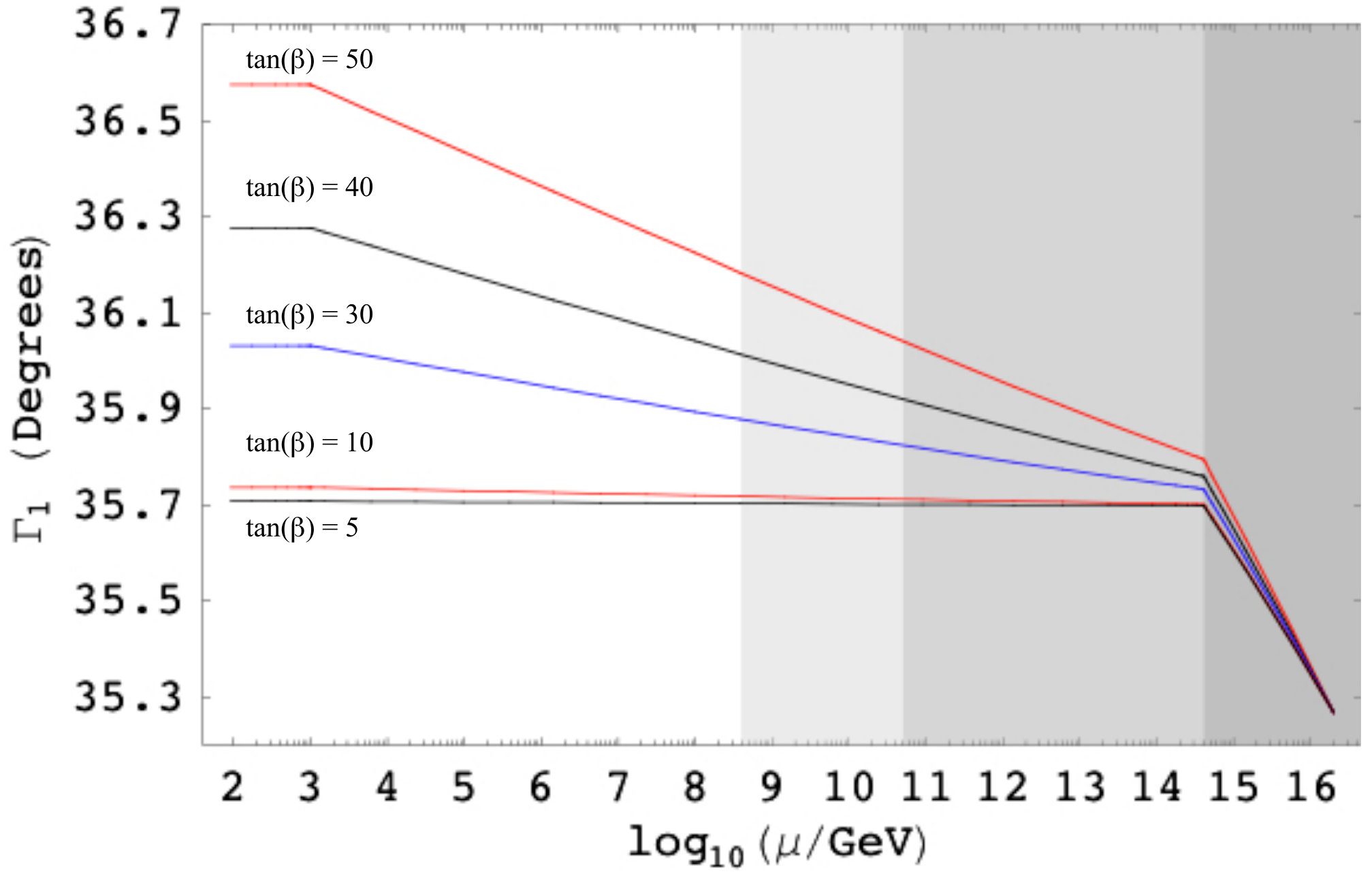}}~
\subfloat[ ]{\label{fig:nm3} \includegraphics[height=56mm,width=81mm]
{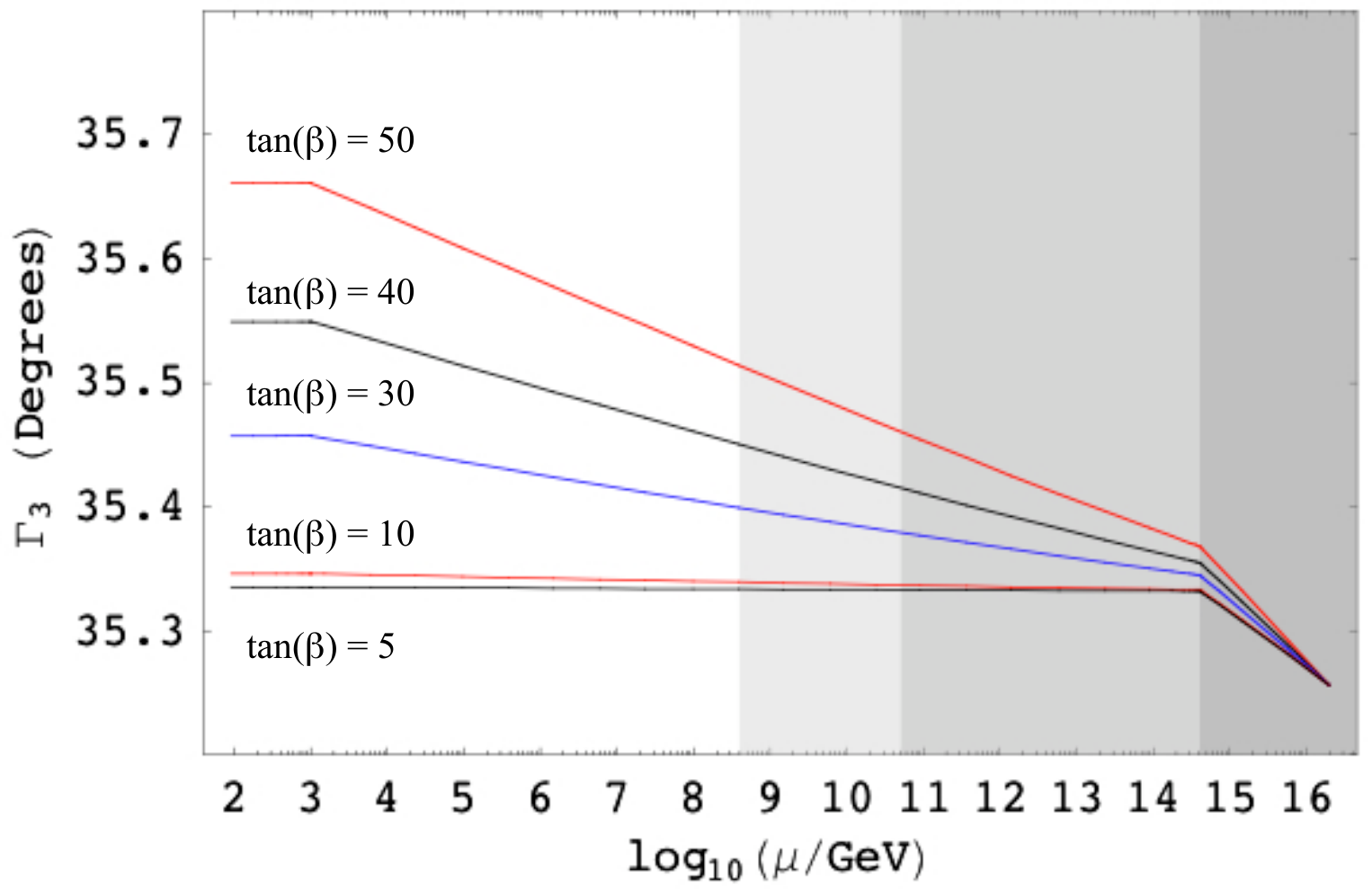}}
\end{center}
\caption{Sum rules with heavy sequential dominance.
This figure shows the evolution of the $\Gamma_i$ sum rules
for $\tan \beta =50$
(panel \protect\subref{fig:nm1}), the running of $\Gamma_1$ and $\Gamma_3$
 in terms of $\tan \beta$
(panels \protect\subref{fig:nm2}, \protect\subref{fig:nm3}) from the GUT scale to the
 electroweak scale
for the case of Cabibbo-like charged lepton corrections with
$\theta_{12}^E = 5 \,^{\circ}$, $\lambda_{12}^E = 0 \,^{\circ}$.}
\label{fig:modc2}
\end{figure}

We assume charged lepton corrections of the form of Eq.\ref{cl}, the neutrino
Yukawa matrix in the non-diagonal charged lepton basis is then transformed
to the diagonal charged lepton basis according to Eq.\ref{rot}. Using the REAP package,
the running of $\Gamma_i$ sum rules was performed from the GUT scale to low energy
 scale and the results are shown in Figure.\ref{fig:modc2}. As shown in this figure,
 despite the larger threshold corrections, for
$\tan(\beta)=50$, the RG running of $\Gamma_3$ is still small (about $0.4^o$)
whereas that of $\Gamma_1$ and $\Gamma_2$ is
about $1.3^o$, compared to the results shown in Figure.\ref{fig:tb}
(nearly $1^o$). This suggests that, qualitatively, the results obtained for the
previous numerical example
inspired by the GUT models in \cite{King:2005bj,deMedeirosVarzielas:2005ax}
are expected to have wide applicability beyond the specific example considered.

\section{Justification of the numerical approach}\label{analytic}

The results in this paper have been based on a numerical
evaluation of the RG corrections using the REAP package. The
reasons why we have chosen to study these corrections numerically,
rather than using analytic estimates \cite{Antusch:2005gp} are as
follows.

The first reason we follow the numerical approach is that, as we
show, some analytic estimates of RG effects which have ignored the
effects of phases are unreliable. For example, the main purpose of
this paper is to find out precisely how large the RG corrections
are to sum rule relations which have been proposed in the
literature (see section 4.1.1). Although the RG corrections to
such sum rules are expected to be small, they are certainly not
negligible compared to the expected precision of future neutrino
experiments, and indeed this prompted the introduction of the
modified sum rule in Eq.\ref{sd2}, where the extra term compared
to Eq.\ref{sd1} was supposed to take into account the RG
corrections \cite{Antusch:2007ib}. However, it turns out that the
new analytic term, which ignores the effects of phases, is too
simplistic. Indeed the numerical results in Figures \ref{fig:sdo}
and \ref{fig:news} clearly show that the extra term included in
the analytic estimate of the RG correction in Eq.\ref{sd2} does
not capture the phase dependence of the RG correction to the
original sum rule in Eq.\ref{sd1}. The numerical study in this
paper has highlighted the shortcoming of analytic estimates of the
RG corrections to sum rules which do not include the phase
dependence.

The second reason we follow the numerical approach, rather than an
analytic approach, is that for some of the cases studied the
analytic approach is simply not applicable. The usual analytic
approach is based on the assumption that only the third family
charged lepton and neutrino Yukawa couplings are taken into
account (while many analytic studies ignore neutrino Yukawa
couplings and threshold effects altogether). Whilst the
approximation of keeping only third family Yukawa couplings is
sufficient for some models, for example the LSD class of models,
it is certainly not sufficient for all classes of models. For
example the HSD case that we also study involves two large
neutrino Yukawa couplings, and the analytic estimates in
\cite{Antusch:2007ib} do not directly apply to this case.

The third reason for following a numerical approach is a purely
quantitative one, namely, even for the cases where the analytic
approach is reliable and applicable (and we have already seen
examples in the previous two paragraphs when it is neither) we
would like to obtain the best possible estimate of the RG
corrections which are the main focus of this paper. If the sum
rules are to be confronted with experiment, it is important to
have a reliable quantitative handle on the RG corrections, and for
this purpose it is necessary to go beyond the leading log analytic
approximation, as we now discuss.

In order to investigate the quantitative accuracy of the analytic
approach, in the remainder of this section we shall compare the
analytic estimates of the RG effects for the LSD example studied
earlier in the paper. For this purpose it is sufficient to switch
off the charged lepton corrections and study the RG corrections to
the neutrino mixing angles using the analytic approximations in
\cite{Antusch:2005gp} which we then compare to the numerical
results we obtained earlier in the paper, and which we also
summarize here for convenience. In order to estimate the RG
corrections to the mixing angles, following \cite{Antusch:2005gp}
it is assumed that the (3,3) matrix elements govern both the
charged Yukawa matrix ($Y^e$) and the neutrino Yukawa matrix
($Y^\nu$)\footnote{We have already noted that for some models such
as HSD this is not the case for the neutrino Yukawa matrix.} in
the flavour basis in which the charged lepton mass matrix is
diagonal. Taking $Y^e \approx diag(0,0,y_\tau)$ and $Y^\nu \approx
diag(0,0,y_{\nu_3})$, one finds, to leading log approximation,
that there is a single parameter which governs the RG corrections
to all the mixing angles given by \cite{Antusch:2007ib}:
\begin{equation}\label{eta}
\eta^{RG}= \frac{y_\tau ^2}{8 \pi^2} \ln \frac{M_{GUT}}{M_Z} +\frac{y_{\nu_3}^2}{8 \pi^2} \ln
\frac{M_{GUT}}{M_3}.
\end{equation}
Assuming tri-bimaximal neutrino mixing at the GUT scale, the low
energy scale parameters are then given approximately by:
\begin{equation}\label{ENp}
s_{12}^\nu (M_Z)=\frac{1}{\sqrt{3}}(1+\frac{\eta^{RG}}{6})~~,~~ s_{23}^\nu (M_Z) = \frac{1}{\sqrt{2}}
(1+\frac{\eta^{RG}}{4})~~,~~s_{13}^\nu (M_Z) =\frac{\eta^{RG}}
{3}\frac{m_2}{m_3}.
\end{equation}

We now apply the above analytic formalism to the LSD model defined
in section.\ref{lsd}, and subsequently studied numerically in this
paper. In this model from Eq.\ref{cor} we see that $y_{\nu_3}= c_3
= 0.58 $ at the GUT scale. We also find $y_\tau =0.33$ and the
mass ratio $m_2/m_3=0.16$ for the case $\tan (\beta) = 50 $. Using
these values, the mixing angles and the quantities ($\Gamma_i$)
can be estimated as shown in Table.\ref{tab7}, where the analytic
estimates are shown alongside the numerical results for
comparison.

 \begin{table}[hbtp]

    \centering

    \begin{tabular}{|l|c|c|c|c|c|c|}

    \hline

    Parameter & $\theta_{12}^\nu $& $\theta_{23}^\nu $& $\theta_{13}^\nu$  & $\Gamma_1$& $
\Gamma_2$& $\Gamma_3$ \\

    \hline

   Analytic ($^o$) & 35.59 & 45.69  &  0.15 & 35.94 & 35.94&35.44  \\

    \hline

  Numerical ($^o$) & 35.65 & 46.00  &  0.15 & 36.21 & 36.21&35.49  \\

    \hline
    \end{tabular}

    \caption{This table shows a comparison between the analytic and numerical results for the RG
corrections to neutrino mixing angles at the $M_Z$ scale, assuming
that they take the precise TB mixing values at the GUT scale, for
the LSD model described in the body of the paper with $\tan
(\beta) = 50 $.}

    \label{tab7}

\end{table}

It is interesting to compare the analytic results to the numerical
results in Table~\ref{tab7} for the neutrino mixing angles at the
$M_Z$ scale, assuming that they take the precise TB mixing values
at the GUT scale and setting all charged lepton corrections to
zero, for the LSD model described above. The results show that the
numerical estimate of $\theta_{13}^\nu$ (which is equal to zero at
the GUT scale) is very accurately reproduced by the analytic
approximation (indeed there is no difference to 2 d.p.), and the
RG correction to $\theta_{12}^\nu$ is also well reproduced with
the analytic estimate underestimating the correction by only
$0.06$ degrees. However the results also show that there is a
significant underestimate of $\theta_{23}^\nu$ with the
analytically estimated value at the $M_Z$ scale being less than
the numerical value by about $0.3$ degrees, resulting in the
analytically estimated values for $\Gamma_1$ and $\Gamma_2$ being
less than the numerical values by about the same amount ($0.3^o$).
From the point of view of the effects studied in this paper (for
example note the precision of the scales shown in the results in
Figure~\ref{fig:sum}) an error of $0.3$ degrees is undesirable and
we would not wish to compromise the results in this paper by being
subject to such unnecessary errors incurred by the analytic
approach.

Finally we remark that the origin of the discrepancy between the
analytic estimates and the numerical results, for the cases where
the analytic approach is reliable and applicable, is due to the
fact that the analytic estimates are based on the assumption that
the Yukawa couplings $y_{\tau}$ and $y_{\nu_3}$ are fixed at their
GUT scale values and do not run, whereas the numerical results
allow for the co-running of all the Yukawa couplings in the matrix
(including the second family Yukawa couplings), with the leading
logs being effectively re-summed.

\section{Summary and Conclusions}

In this paper we have analyzed the effects of charged lepton
corrections and RG running on the low energy predictions of
theories which accurately predict tri-bimaximal neutrino mixing at
the high energy scale. In GUT motivated examples the charged
lepton corrections are often Cabibbo-like and in this case the
effect of charged lepton corrections leads to a range of neutrino
mixing sum rules at the GUT scale, given by the $\Gamma_i$ sum
rules in Eqs.\ref{sm},\ref{su},\ref{gamma3}, as well as the
$\sigma_i$ sum rules expressed in terms of the deviation
parameters in Eqs.\ref{sd1} and \ref{sd2}. We have studied the RG
running of such sum rules numerically for a specific numerical
example inspired by the GUT models in
\cite{King:2005bj, deMedeirosVarzielas:2005ax}, corresponding
closely to CSD with LSD. Our results indicate small but measurable
effects for the two examples studied. For example the $\Gamma_3$
sum rule arising from Cabibbo-like charged lepton corrections (due
to $\theta_{12}^E$ corrections) which at the GUT scale corresponds
to $\theta_{12} -\theta_{13} \cos (\delta) \approx 35.3^o$ becomes
renormalized by about $0.4^o$ even for large $\tan \beta=50$. We
have also considered the effect on charged lepton corrections
coming from non-Cabibbo-like charged lepton corrections (due to
non-zero $\theta_{23}^E$) and confirmed that the sum rule
$\sigma_1$ is insensitive to $\theta_{23}^E$.

Even for a particular class of numerical model, such as the
GUT-flavour inspired LSD model considered,
the numerical results will depend in general on the choice of Majorana phases for that model.
We have seen that switching on these Majorana phases can alter significantly the running of the
TB mixing deviation parameters $r,s,a$ as well as the sum rules such as $\sigma_i$.
For example the sum rule $\sigma_2$ which includes the leading logarithmic RG corrections
due to the running of $s$ and $a$, will have a Majorana phase dependence via the running of $r$
which was neglected in the derivation of $\sigma_2$ \cite{Antusch:2007ib}. Thus, the relative
stability of $\sigma_2$
as compared to $\sigma_1$ turns out to be a Majorana phase dependent question.

Although most of the numerical results are based on a particular GUT-flavour motivated LSD type
of model,
we have also considered similar results for a completely different type of model
based on HSD. Overall we have found that the
RG running effects are quite small in both cases which suggests that qualitatively
similar results will apply to other models based on the Minimal Supersymmetric
Standard Model, extended to include the see-saw mechanism,
with hierarchical neutrino masses. However, we repeat our caveat in the Introduction that
in this paper we have only considered deviations from TB lepton mixing due to the
combination of charged lepton corrections
and RG running, and that in general there will be other sources of deviations
that we do not consider. For example, exact CSD \cite{King:2005bj} itself will lead to some
deviations since it does not predict precisely TB neutrino mixing
due to corrections of order $m_2/m_3$ \cite{King:1998jw}.
Another example of deviations to TB mixing are the
canonical normalization effects discussed in \cite{Antusch:2007ib}.
In general corrections to the TB mixing in the
neutrino sector, in the flavour basis, are highly model
dependent and have been studied phenomenologically
in \cite{Albright}.

Finally we note that all the results in this paper are based on a
numerical approach rather than an analytic approach. We have given
three reasons to justify the use of the numerical approach:
firstly that analytic results which ignore the effects of phases
are not reliable (for example the $\sigma_2$ sum rule); secondly
that analytic results may not be applicable for all cases of
interest (for example HSD with two large neutrino Yukawa
couplings); and thirdly that the quantitative precision of the
analytic approach is not in general sufficient for our purpose
here (in particular we have shown that the RG correction to
$\theta_{23}$ may be significantly underestimated in the analytic
approach). Therefore throughout this paper we have adopted a
numerical approach, exploiting the convenient REAP package.

To conclude, we have assumed precise TB neutrino mixing at the GUT
scale, and investigated the deviations due to both charged lepton
corrections and RG running. In GUT-flavour inspired models the
charged lepton corrections are expected to resemble those of the
quark mixing matrix, and lead to neutrino mixing sum rules which
are valid at high energy. We have studied numerically a variety of
such neutrino mixing sum rules and shown that, in the types of
GUT-flavour models in which the sum rules emerge in the first
place, they are subject to only mild RG corrections (less than one
degree for the cases studied) in evolving them down to low energy.
However even such small corrections as described in this paper
will nevertheless be important when comparing the neutrino mixing
sum rules to the results of future high precision neutrino
oscillation experiments \cite{Bandyopadhyay:2007kx}.

\section*{Acknowledgements}
We are deeply indebted to Stefan Antusch for his many helpful suggestions and
comments throughout this work.
We also acknowledge partial support from the following grants:
PPARC Rolling Grant PPA/G/S/2003/00096;
EU Network MRTN-CT-2004-503369;
EU ILIAS RII3-CT-2004-506222.

\end{document}